%% file: VortexLieTransformPP.tex
\documentclass[aps, onecolumn]{revtex4-1}
\usepackage[dvips]{graphicx}
\usepackage{color}
\usepackage{amssymb,amsmath,amsfonts}

\newcommand{\mybold}[1]{{\mbox{\bf\boldmath ${#1}$}}}
\newcommand{\hodge}[1]{{{}^*{#1}}}

\newcommand{\bfr}{{\mybold{r}}}
\newcommand{\bfR}{{\mybold{R}}}
\newcommand{\bfv}{{\mybold{v}}}
\newcommand{\bfw}{{\mybold{w}}}
\newcommand{\bfz}{{\mybold{z}}}

\newcommand{\mybar}[1]{{\overline{{#1}}}}
\newcommand{\bracketnewln}[1]{\right.\\#1&\left.{}}

\begin{document}

\title{A Robust Numerical Method for Integration of\\ Point-Vortex Trajectories in Two Dimensions}

\author{Spencer A. Smith}
\author{Bruce M. Boghosian}

\affiliation{Department of Physics and Department of Mathematics, Tufts University, Medford, Massachusetts 02420, USA}

\date{\today}

\begin{abstract}

The venerable 2D point-vortex model plays an important role as a simplified version of many disparate physical systems, including superfluids, Bose-Einstein condensates, certain plasma configurations, and inviscid turbulence.  This system is also a veritable mathematical playground, touching upon many different disciplines from topology to dynamic systems theory.  Point-vortex dynamics are described by a relatively simple system of nonlinear ODEs which can easily be integrated numerically using an appropriate adaptive time stepping method.  As the separation between a pair of vortices relative to all other inter-vortex length scales decreases, however, the computational time required diverges.  Accuracy is usually the most discouraging casualty when trying to account for such vortex motion, though the varying energy of this ostensibly Hamiltonian system is a potentially more serious problem.  We solve these problems by a series of coordinate transformations:  We first transform to action-angle coordinates, which, to lowest order, treat the close pair as a single vortex amongst all others with an internal degree of freedom.  We next, and most importantly, apply Lie transform perturbation theory to remove the higher-order correction terms in succession.  The overall transformation drastically increases the numerical efficiency and ensures that the total energy remains constant to high accuracy. 

\end{abstract}


\maketitle

\newpage

\tableofcontents

\newpage

\section{Introduction}

\subsection{Importance of Point-Vortex Dynamics}

	Despite its relative simplicity, the 2D point-vortex model has appeared time and again in the description of a wide variety of physical systems.  Part of this breadth of application is attributable to its long history starting with Helmholtz in 1858~\cite{bib:Aref2}, who considered point-like distributions of vorticity imbedded in a 2D ideal, incompressible fluid.  Indeed, the velocity field generated by the Hamiltonian motion of a collection of point vortices satisfies the Euler fluid equation~\cite{bib:saffman}.  This represents a remarkable computational and conceptual simplification: trading the infinite dimensional ideal fluid field equations for the finite dimensional coupled ordinary differential equations of the point-vortex model.  It is completely natural then to ask: to what extent are various general phenomena in 2D hydrodynamics present in the point-vortex model? \\

	Onsager was one of the first to tackle this quiestion when he used ideas from equilibrium statistical mechanics to show the existence of negative temperature states~\cite{bib:EyinkSpohn}; states that correspond to large-scale, long-lived vortex structures, not unlike those that form in Earth's atmosphere.  Further research in the statistical vein has resulted in a kinetic theory of point vortices~\cite{bib:Chavanis}, while consideration of structure formation dovetails nicely with the concept of Lagrangian coherent structures~\cite{bib:HallerYuan}~\cite{bib:Provenzale}.  There is also the view that point vortices are a useful toy model of 2D inviscid turbulence~\cite{bib:Babiano}~\cite{bib:Tabeling}~\cite{bib:Benzi}, an idea motivated by the non-integrable, i.e., chaotic, motion of four or more vortices.  Closely related to this is the more rigorous concept of chaotic advection~\cite{bib:Aref1}, which can help explain the transport properties of a vortex dominated ideal fluid~\cite{bib:Leoncini}.  Clearly a wide variety of physical phenomena falls under the aegis of point-vortex dynamics. \\
	
	In addition to a diverse spectrum of phenomena in ideal fluids, the point-vortex model is also applicable to other, more exotic, physical systems.   This includes 2D electron plasmas~\cite{bib:Driscoll1}.  Indeed, it was in the plasma physics community where Lie perturbation theory was first used with great success~\cite{bib:Littlejohn}~\cite{bib:Cary2}.  Bose-Einstein Condensates (BECs) also fall into this category.  The Gross-Pitevskii equation, which governs the evolution of the BEC wave-function, can be re-expressed, via the Madelung transformation, as Euler's equations~\cite{bib:JerrardSpirn}.  Therefore, quantized vortex defects in a rotating BEC~\cite{bib:Fetter}~\cite{bib:WeilerNeely} interact, to first approximation, as if they are point vortices.  This also applies to other superfluids such as He-II, where the vortex circulations are still quantized and the size of the vortex core is small enough to really warrant approximation by point vortices.  Since superfluid turbulence is dictated by 3D quantized line vortices~\cite{bib:Vinen}, the point-vortex model can be considered a toy model of quantum turbulence as well.\\
	
	Aside from physical instantiations, the point-vortex model should engender some intrinsic interest simply as an interesting mathematical entity.  It has been described as a mathematical playground~\cite{bib:Aref2}, touching upon areas such as the theory of dynamic systems, ODEs, and Hamiltonian dynamics, whose appearance might be expected, as well as some ideas that at first seem to have no connection.  For example, neither is it readily apparent that equilibrium configurations of point vortices can be connected to the roots of certain polynomials~\cite{bib:Aref3}, nor is it immediate that one can apply topology in the guise of Nielsen-Thurston and braid theory to describe fluid mixing~\cite{bib:BoylandStremler}.  A plethora of physical and mathematical considerations give weight to the notion that the point-vortex system is an important item of study despite its relative simplicity.  \\

\subsection{Posing the Problem}

	Orbits of the point-vortex dynamical system can be obtained easily by numerically integrating the set of coupled ODEs with an appropriate adaptive time-stepping method.  When two like-signed vortices approach each other, they simply rotate about their center of circulation with an angular frequency that is inversely dependent on the separation length, squared.  Therefore, in a system of many vortices, the closer a vortex pair is, the faster their angular movement will be compared to that of other vortices.  To accommodate this motion, the integration method will reduce the time-steps between integrator function calls to maintain the prescribed tolerance.  This prohibitively and unnecessarily slows down the integration of the system as a whole.  One naive way of circumventing this problem is to adaptively change the tolerance, though this has the unpalatable consequence of decreasing the accuracy of the computed orbit.  While one could possibly ignore this with appeals to the chaotic, non-integrable nature of the dynamics, there is another serious problem to consider.  In most integrators the decreased accuracy will cause the vortices to systematically overshoot their ideal near-circular orbit.  This has the effect of increasing the vortex-pair separation, and therefore decreasing the energy in this ostensibly Hamiltonian system.  It turns out that one does not have to choose either lowered accuracy and changing energy or long integration times.  We develop a transformation in this paper which largely alleviates both of these problems, and does so in a physically intriguing way.

\subsection{Outline of the Solution}

	Intuitively, a pair of like-signed vortices that are considered close when compared to all other inter-vortex distances -- what we shall call a vortex ``dimer" -- will rotate about its center in a manner that appears unaffected by the presence of all other vortices.  That is, the internal motion of the dimer, that of the constituent vortex pairs with respect to their center of circulation, is approximately the integrable motion of the pair on their own.  Similarly, other vortices will interact with the dimer roughly as if it is a single vortex.  This crude approximation has the useful feature that the now integrable dimer motion ceases to be the limiting factor in choosing the step size for numerical integration.  \\
	
	Of course, we can not perfectly shoe-horn our problem into this picture.  After some initial transformations to express the dimer in action-angle coordinates, we find correction terms which modify this picture.  Fortunately these terms can be expanded in powers of a small factor, and therefore our system becomes amenable to perturbation methods.  In particular, we use Lie transform perturbation theory, see ~\cite{bib:DragtFinn}~\cite{bib:Deprit}~\cite{bib:Cary1}~\cite{bib:Littlejohn2}, because the invariance of the symplectic structure of our phase space is manifest with these transforms.  From here, we choose generating functions for the Lie transforms that will get rid of the correction terms successively at each order, while maintaining, as best we can, the integrable nature of the dimer's internal motion.  When a correction term which we can not transform away does arise, we will see that it does not present any real computational problem.  Indeed it will even have the interesting physical interpretation of an additional field that couples only to vortices which have an internal ``spin" degree of freedom, i.e., the dimers.  Overall we will have succeeded in converting the original N-body problem into an (N-1)-body problem.\\
	
	We will show that the overall transformation alleviates the small time-step issue, maintains a constant energy, and results in very accurate orbits.  It should be noted that KAM theory and other superconvergent methods are not applicable here, seeing that they can not deal with the non-integrable system which arises after our first Lie transformation.  Likewise it should be clear that our method is not connected with fast multipole methods, which increase the speed of integrating an N-body system by including only the most important pairwise interactions for each particle.  We consider all $N(N-1)/2$ pairwise calculations, and therefore do not attempt to change how the algorithmic complexity scales with point-vortex number.  We are more concerned with accuracy and how we can preserve it, while solving the specific small time-step issue.

\section{Point-Vortex Dynamics}
\subsection{Equations of Motion}

We consider a system of $N$ point vortices in two spatial dimensions, each having position $\bfr_j=\langle x_j, y_j\rangle$ and circulation $\Gamma_j$.  Their dynamics is described by the equations of motion,
\begin{equation}
\frac{d\bfr_j}{dt} = \sum_{k\neq j}^N\frac{\Gamma_k}{2\pi}\;\frac{\hodge{(\bfr_j-\bfr_k)}}{|\bfr_j-\bfr_k|^2}.  \label{eq:InitialDynamics}
\end{equation}
Here we have defined the $\hodge{}$ operator, so that if $\bfv=\langle v_x,v_y\rangle$ is a vector in two dimensions, then $\hodge{\bfv}\equiv\langle -v_y,v_x\rangle$, i.e., a rotation by $\frac{\pi}{2}$.  These equations of motion can then be decomposed into coordinates,
\begin{eqnarray}
\frac{dx_j}{dt} &=& -\sum_{k\neq j}^N\frac{\Gamma_k}{2\pi}\;\frac{y_j-y_k}
{(x_j-x_k)^2+(y_j-y_k)^2}\label{eq:xvelocity}\\
\frac{dy_j}{dt} &=& +\sum_{k\neq j}^N\frac{\Gamma_k}{2\pi}\;\frac{x_j-x_k}
{(x_j-x_k)^2+(y_j-y_k)^2}.\label{eq:yvelocity}
\end{eqnarray}
For reasons that will become clear, the $2N$-dimensional space whose coordinates are $\bfr_1,\ldots,\bfr_N$ will be called the {\it phase space} of the vortices.  The state of all $N$ vortices in the system is represented by a single point in phase space.

\subsection{Hamiltonian Formulation}

The scalar function on phase space,
\begin{equation}
H\left(\bfr_1,\ldots,\bfr_N\right)\equiv
-\sum_{j}^N\sum_{k\neq j}^N
\frac{\Gamma_j\Gamma_k}{4\pi}\ln \left|\bfr_j-\bfr_k\right|.
\label{eq:hamfun}
\end{equation}
is called the {\it Hamiltonian}.  It is easily verified that Eqs.~(\ref{eq:xvelocity}) and (\ref{eq:yvelocity}) can be rewritten in terms of the Hamiltonian:

\begin{eqnarray}
\frac{dx_j}{dt} &=& +\frac{1}{\Gamma_j}\;\frac{\partial H}{\partial y_j}
\label{eq:canonicalx}\\
\frac{dy_j}{dt} &=& -\frac{1}{\Gamma_j}\;\frac{\partial H}{\partial x_j}
\label{eq:canonicaly}
\end{eqnarray}
Apart from the factors of $1/\Gamma_j$, these have the form of Hamilton's canonical equations of motion, where $x_j$ and $y_j$ are canonically conjugate variables.

Eqs.~(\ref{eq:canonicalx}) and (\ref{eq:canonicaly}) are rather unusual, in that most examples of Hamiltonian systems pair position coordinates with canonically conjugate momentum coordinates.  This system, by contrast, has pairs of canonically conjugate position coordinates.  To pursue this identification, define the {\it Poisson bracket} of two functions
\begin{equation}
\left\{ A,B \right\} =
\sum_j^N\frac{1}{\Gamma_j}\left(
\frac{\partial A}{\partial x_j}\frac{\partial B}{\partial y_j} -
\frac{\partial A}{\partial y_j}\frac{\partial B}{\partial x_j}
\right).
\label{eq:pb}
\end{equation}
It is readily verified that this bracket maps any two smooth ($C^\infty$) functions of $\bfr_1,\ldots,\bfr_N$ to a third, is bilinear
\begin{equation}
\left\{\alpha A + \beta B, C\right\} = \alpha \left\{A,C\right\} + \beta\left\{B,C\right\}
\label{eq:bilinear}
\end{equation}
(where $\alpha$ and $\beta$ are numbers), antisymmetric
\begin{equation}
\left\{B,A\right\} = -\left\{A,B\right\},
\label{eq:antisymmetric}
\end{equation}
and obeys the Jacobi identity
\begin{equation}
\left\{\left\{A,B\right\},C\right\} + \left\{\left\{C,A\right\},B\right\} + \left\{\left\{B,C\right\},A\right\} = 0.
\label{eq:jacobi}
\end{equation}

In particular, we see that the phase-space coordinates themselves satisfy the bracket relations
\begin{eqnarray}
\left\{ x_j, x_k \right\} = \left\{ y_j, y_k \right\} &=& 0\\
\left\{ x_j, y_k \right\} &=& \delta_{jk}/\Gamma_j.
\end{eqnarray}
Technically, because of the factor of $1/\Gamma_j$, the Poisson bracket defined by Eq.~(\ref{eq:pb}) is {\it noncanonical}.  It is possible to restore a canonical bracket by scaling the coordinates of the $j$'th vortex by factors of $\sqrt{|\Gamma_j|}$, swapping $x_j$ and $y_j$ if $\Gamma_j$ is negative~\cite{bib:saffman}, but it seems easier to simply live with the noncanonical bracket that has emerged naturally from this analysis.  All that is really required of a Poisson bracket is that it satisfy Eqs.~(\ref{eq:bilinear}), (\ref{eq:antisymmetric}) and (\ref{eq:jacobi}), and our noncanonical bracket does do so.

In terms of the Poisson bracket, the canonical equations of motion may be written
\begin{eqnarray}
\frac{dx_j}{dt} &=& \left\{x_j, H\right\}\\
\frac{dy_j}{dt} &=& \left\{y_j, H\right\}.
\end{eqnarray}
Indeed, the rate of change of any function on phase space, $A(\bfr_1,\ldots,\bfr_N)$, is given by
\begin{equation}
\frac{dA}{dt}
=
\sum_j^N\left(
\frac{\partial A}{\partial x_j}\frac{dx_j}{dt} + 
\frac{\partial A}{\partial y_j}\frac{dy_j}{dt}
\right)
=
\sum_j^N\frac{1}{\Gamma_j}\left(
\frac{\partial A}{\partial x_j}\frac{\partial H}{\partial y_j} -
\frac{\partial A}{\partial y_j}\frac{\partial H}{\partial x_j}
\right)
=
\left\{ A,H \right\}.
\end{equation}

Conservation of energy is then established by noting that
\begin{equation}
\frac{dH}{dt} = \left\{ H,H \right\} = 0,
\label{eq:EnergyConservation}
\end{equation}
which follows from the antisymmetry of the bracket.  That is, the Hamiltonian is constant, and numerically equal to the vortices' energy,
\begin{equation}
H\left(\bfr_1,\ldots,\bfr_N\right)=E,
\label{eq:energyconstraint}
\end{equation}
constraining the dynamics to lie on a surface of codimension one in phase space.

\section{A Vortex Dimer}
\subsection{Hamiltonian for a Vortex Dimer}

Returning to Eq.~(\ref{eq:hamfun}) for the Hamiltonian of $N$ vortices in an infinite domain, we suppose that two like-signed vortices are very close to one another.  We further suppose that these two are the ones labeled $m$ and $n$, and we refer to this pair as a ``vortex dimer.''  We rewrite the Hamiltonian as follows:
\begin{eqnarray}
H\left(\bfr_1,\ldots,\bfr_m,\ldots,\bfr_n,\ldots,\bfr_N\right)
&\equiv&
-\frac{\Gamma_m\Gamma_n}{2\pi}\ln \left|\bfr_m-\bfr_n\right|
-\sum_{j\neq m,n}^N\sum_{k\neq j,m,n}^N\frac{\Gamma_j\Gamma_k}{4\pi}\ln \left|\bfr_j-\bfr_k\right|
\nonumber\\
& &
-\sum_{j\neq m,n}^N\frac{\Gamma_m\Gamma_j}{2\pi}\ln \left|\bfr_m-\bfr_j\right|
-\sum_{j\neq m,n}^N\frac{\Gamma_n\Gamma_j}{2\pi}\ln \left|\bfr_n-\bfr_j\right|.
\end{eqnarray}
The first term on the right is the Hamiltonian of interaction between vortices $m$ and $n$ within the dimer.  The second term is the Hamiltonian of interaction amongst all of the other vortices in the system.  The third and fourth terms describe the interaction of the dimer with all the other vortices in the system; in particular, the third term describes the interaction between all of the other vortices and vortex $m$, and the fourth term describes the interaction between all of the other vortices and vortex $n$.

\subsection{Transformation to Center of Circulation and Relative Coordinates}

We introduce the {\it center of circulation} of the vortex dimer
\begin{equation}
\bfR\equiv\frac{\Gamma_m\bfr_m+\Gamma_n\bfr_n}{\Gamma_m+\Gamma_n},
\end{equation}
and its {\it relative displacement}
\begin{equation}
\bfr\equiv\bfr_n - \bfr_m.
\end{equation}
We transform coordinates to eliminate $\bfr_m$ and $\bfr_n$ in favor of $\bfR=\langle X,Y\rangle$ and $\bfr=\langle x,y\rangle$.  The inverse transformation is
\begin{eqnarray}
\bfr_m &=& \bfR - \frac{\Gamma_n}{\Gamma_m + \Gamma_n}\bfr\label{eq:inversem}\\
\bfr_n &=& \bfR + \frac{\Gamma_m}{\Gamma_m + \Gamma_n}\bfr.\label{eq:inversen}
\end{eqnarray}
The Poisson bracket relations among the new coordinates may readily calculated to be:
\begin{equation}
\left\{ X, Y \right\} = \frac{1}{\Gamma_R},
\end{equation}
\begin{equation}
\left\{ x, y \right\} = \frac{1}{\Gamma_r},
\end{equation}
 and
\begin{equation}
\{x, Y\} = \{y, X\} = 0,
\end{equation}
where the {\it total circulation} of the vortex dimer is
\begin{equation}
\Gamma_R\equiv\Gamma_m + \Gamma_n,
\end{equation}
and the {\it reduced circulation} of the vortex dimer is
\begin{equation}
\Gamma_r\equiv\frac{\Gamma_m\Gamma_n}{\Gamma_m + \Gamma_n}.
\end{equation}
It follows that the Poisson bracket in the new coordinates is
\begin{equation}
\left\{ A, B\right\} \equiv
\frac{1}{\Gamma_R}\left(\frac{\partial A}{\partial X}\frac{\partial B}{\partial Y} - 
\frac{\partial A}{\partial Y}\frac{\partial B}{\partial X}\right) +
\frac{1}{\Gamma_r}\left(\frac{\partial A}{\partial x}\frac{\partial B}{\partial y} - 
\frac{\partial A}{\partial y}\frac{\partial B}{\partial x}\right) +
\sum_{j\neq m, n}\frac{1}{\Gamma_j}
\left(\frac{\partial A}{\partial x_j}\frac{\partial B}{\partial y_j} - 
\frac{\partial A}{\partial y_j}\frac{\partial B}{\partial x_j}\right).
\end{equation}
We note that $X$ and $Y$ comprise a canonically conjugate pair, as do $x$ and $y$.

It remains to write the Hamiltonian in the new coordinates.  Using Eqs.~(\ref{eq:inversem}) and (\ref{eq:inversen}), this is straightforward, and we find
\begin{eqnarray}
H
&=&
-\frac{\Gamma_R\Gamma_r}{2\pi}\ln\left|\bfr\right|
-\sum_{j\neq m,n}^N\sum_{k\neq j,m,n}^N\frac{\Gamma_j\Gamma_k}{4\pi}\ln \left|\bfr_j-\bfr_k\right|
\nonumber\\
& &
-\sum_{j\neq m,n}^N\frac{\Gamma_m\Gamma_j}{2\pi}
\ln \left|\bfR-\bfr_j - \frac{\Gamma_r}{\Gamma_m}\bfr\right|
-\sum_{j\neq m,n}^N\frac{\Gamma_n\Gamma_j}{2\pi}
\ln \left|\bfR-\bfr_j + \frac{\Gamma_r}{\Gamma_n}\bfr\right|
\nonumber\\
&=&
-\frac{\Gamma_R\Gamma_r}{2\pi}\ln\left|\bfr\right|
-\sum_{j\neq m,n}^N\sum_{k\neq j,m,n}^N\frac{\Gamma_j\Gamma_k}{4\pi}\ln \left|\bfr_j-\bfr_k\right|
\nonumber\\
& &
-\sum_{j\neq m,n}^N\frac{\Gamma_m\Gamma_j}{4\pi}
\ln \left[
\left|\bfR-\bfr_j\right|^2 - 2\frac{\Gamma_r}{\Gamma_m}\left(\bfR-\bfr_j\right)\cdot\bfr + \frac{\Gamma_r^2}{\Gamma_m^2}r^2\right]
\nonumber\\
& &
-\sum_{j\neq m,n}^N\frac{\Gamma_n\Gamma_j}{4\pi}
\ln \left[
\left|\bfR-\bfr_j\right|^2 + 2\frac{\Gamma_r}{\Gamma_n}\left(\bfR-\bfr_j\right)\cdot\bfr + \frac{\Gamma_r^2}{\Gamma_n^2}r^2\right]
\nonumber\\
&=&
-\frac{\Gamma_R\Gamma_r}{2\pi}\ln\left|\bfr\right|
-\sum_{j\neq m,n}^N\sum_{k\neq j,m,n}^N\frac{\Gamma_j\Gamma_k}{4\pi}\ln \left|\bfr_j-\bfr_k\right|
-\sum_{j\neq m,n}^N\frac{\Gamma_R\Gamma_j}{2\pi}\ln\left|\bfR-\bfr_j\right|
\nonumber\\
& &
-\sum_{j\neq m,n}^N\frac{\Gamma_m\Gamma_j}{4\pi}
\ln \left(1 - 2\frac{\Gamma_r}{\Gamma_m}\frac{\left(\bfR-\bfr_j\right)\cdot\bfr}{\left|\bfR-\bfr_j\right|^2} + 
\frac{\Gamma_r^2}{\Gamma_m^2}\frac{|\bfr|^2}{\left|\bfR-\bfr_j\right|^2}\right)
\nonumber\\
& &
-\sum_{j\neq m,n}^N\frac{\Gamma_n\Gamma_j}{4\pi}
\ln \left(1 + 2\frac{\Gamma_r}{\Gamma_n}\frac{\left(\bfR-\bfr_j\right)\cdot\bfr}{\left|\bfR-\bfr_j\right|^2} + 
\frac{\Gamma_r^2}{\Gamma_n^2}\frac{|\bfr|^2}{\left|\bfR-\bfr_j\right|^2}\right).
\label{eq:comhamiltonian}
\end{eqnarray}

As we will be referring to the terms of this Hamiltonian often, it makes sense to label them.  The first three terms are collectively denoted $H_0$.  The first term on its own is $H_{01}$, while $H_{02}$ corresponds to the second and third terms together.  The Hamiltonian of interaction between the two vortices within the dimer is $H_{01}$.  The first term of $H_{02}$ is the interaction Hamiltonian of all non-dimer vortices in the system, and the second term of $H_{02}$ is the principal interaction Hamiltonian between the dimer and all of the other vortices.  Altogether $H_0$ makes the crude approximation that the dimer, while having internal degrees of freedom, is merely a vortex of circulation $\Gamma_R$ at position $\bfR$, interacting normally with the N-2 other vortices.  The rest of the Hamiltonian constitutes correction terms to this picture.

\section{Ordering}
\subsection{Ordering the Hamiltonian}

We would like to develop an ordering scheme whereby the fastest and most important motion is the oscillation of the relative displacement vector of the vortex dimer, described by the first term $\left(H_{01}\right)$ in Eq.~(\ref{eq:comhamiltonian}).   We introduce a formal ordering parameter, $\epsilon$, where the small quantities, $|\bfr|/|\bfR-\bfr_j|$ and $|\bfr|/|\bfr_i-\bfr_j|$, are of order $\epsilon$ or smaller.  Eq.~(\ref{eq:comhamiltonian}) is thus rewritten
\begin{eqnarray}
H
&=&
-\frac{\Gamma_R\Gamma_r}{2\pi}\ln\left|\bfr\right|
-\sum_{j\neq m,n}^N\sum_{k\neq j,m,n}^N\frac{\Gamma_j\Gamma_k}{4\pi}\ln \left|\bfr_j-\bfr_k\right|
-\sum_{j\neq m,n}^N\frac{\Gamma_R\Gamma_j}{2\pi}\ln\left|\bfR-\bfr_j\right|
\nonumber\\
& &
-\sum_{j\neq m,n}^N\frac{\Gamma_m\Gamma_j}{4\pi}
\ln \left(1 - 2\epsilon\frac{\Gamma_r}{\Gamma_m}\frac{\left(\bfR-\bfr_j\right)\cdot\bfr}{\left|\bfR-\bfr_j\right|^2} + 
\epsilon^2\frac{\Gamma_r^2}{\Gamma_m^2}\frac{|\bfr|^2}{\left|\bfR-\bfr_j\right|^2}\right)
\nonumber\\
& &
-\sum_{j\neq m,n}^N\frac{\Gamma_n\Gamma_j}{4\pi}
\ln \left(1 + 2\epsilon\frac{\Gamma_r}{\Gamma_n}\frac{\left(\bfR-\bfr_j\right)\cdot\bfr}{\left|\bfR-\bfr_j\right|^2} + 
\epsilon^2\frac{\Gamma_r^2}{\Gamma_n^2}\frac{|\bfr|^2}{\left|\bfR-\bfr_j\right|^2}\right)
\label{eq:orderedhamiltonian}
\end{eqnarray}

One should immediately note that in this scheme, all of $H_0$ is of order zero in $\epsilon,$ and therefore the internal dimer motion alone does not yet enjoy the distinction of being the unperturbed motion.  We will soon remedy this, when we also order the Poisson bracket, or equivalently the Poisson tensor, in $\epsilon$.  The Hamiltonian ordering does, however, indicate that the correction terms are of order $\epsilon^2$ and higher, though it is not yet manifest.

\subsection{Poisson Tensor}

Lie transform perturbation theory will require that we use the Lie derivative, see~Eq.(\ref{eq:LieDerivativeScalar}).  To ensure that the Lie derivative will deal with the $\epsilon$ ordering correctly, we will need to order the Poisson bracket as well as the Hamiltonian.  For the sake of transparency we adopt the Poisson tensor framework of symplectic geometry instead of the Poisson bracket formalism.  These are both interchangeable, but ordering, and indeed perturbing, the Poisson tensor is conceptually much simpler.  We can define can define the Poisson tensor $\mathbb{P}$ in terms of the Poisson bracket:
\begin{equation}
\mathbb{P}^{jk} = \left\{ z^j, z^k\right\}.
\end{equation}
This Poisson tensor has constant components, see~Eq.(\ref{eq:PoissonTensor}) below.  With it we can rewrite the equations of motion as
\begin{equation}
\dot{z}^j = \mathbb{P}^{jk}\frac{\partial H}{\partial z^k},
\end{equation}
which we will usually refer to in the following equivalent shorthand notation
\begin{equation}
\dot{z} = \mathbb{P}dH.
\end{equation}

\subsection{Ordering the Poisson Tensor}

We have two goals that we wish to accomplish with the ordering of the Poisson tensor.  The first goal, alluded to in the previous section, deals with how the Lie derivative handles the ordering.  This primary concern will be explained later, after we have introduced the Lie derivative machinery.  However, we will state here the ordering of the Poisson tensor that arises from this consideration:
\begin{equation}
\mathbb{P} =  \mathbb{P}_0 + \epsilon^2 \mathbb{P}_2
\label{eq:PoissonOrdering}
\end{equation}
where the ${2\choose 0}$-tensors $\mathbb{P}_0$ and $\mathbb{P}_2,$ expressed as $2N \times 2N$ matrices, are:
\begin{equation}
 \mathbb{P}_0 = 
 \begin{pmatrix}  0 & \mathbb{A} \\
  -\mathbb{A} & 0
 \end{pmatrix}, \;
 \mathbb{P}_2 =  
 \begin{pmatrix}  0 & \mathbb{B} \\
  -\mathbb{B} & 0
 \end{pmatrix}, \; \; \text{where} \; \;
 \mathbb{A} = 
 \begin{pmatrix} 
  \frac{1}{\Gamma_r} &  & \text{\huge{0}} \\
   & 0 &  \\
  \text{\huge{0}} & & \ddots
 \end{pmatrix}, \; \; \text{and} \; \;
 \mathbb{B} = 
 \begin{pmatrix} 
  0 &  &  & \text{\huge{0}}  \\
   & \frac{1}{\Gamma_R} & & \\
   & & \frac{1}{\Gamma_{r_j}} & \\
   \text{\huge{0}} & &  & \ddots 
 \end{pmatrix}.
 \label{eq:PoissonTensor}
 \end{equation}
Our second goal is to recognize that the fast internal movement of the dimer is the most important motion, and therefore elevate it to the status of unperturbed motion in our eventual Lie transform perturbation theory.  The Poisson tensor ordering of~Eq.(\ref{eq:PoissonOrdering}) achieves this.  That is, the zero-order term of $\dot{z} = \mathbb{P}dH,$ when expanded in $\epsilon,$ is the motion of the dimer alone:
\begin{eqnarray}
\dot{z} &=& \left(\mathbb{P}_0 + \epsilon^2 \mathbb{P}_2\right)d\left(H_0 + \mathcal{O}(\epsilon^{2})\right) \; = \; \mathbb{P}_0 dH_{01} + \epsilon^2 \mathbb{P}_2 dH_{02} +  \mathcal{O}(\epsilon^{2}), \;  \; or \\
\dot{z}^j &=& \mathbb{P}_0^{jk} \frac{\partial H_{01}}{\partial z^k} + \epsilon^2 \mathbb{P}_2^{jk} \frac{\partial H_{02}}{\partial z^k} +  \mathcal{O}(\epsilon^{2}).
\end{eqnarray}

While this justifies our consideration of the internal dimer motion as the unperturbed motion, one should take care in interpreting the rest of the above dynamic expansion.  In particular, the ordering is absolutely correct only when comparing the relative size of different contributions to the evolution of an individual coordinate, and not when comparing a single contribution to the evolution of one coordinate with that of another coordinate.  This is a minor issue, and certainly will not affect the development of our overall Lie transform method.

\section{The Unperturbed Problem}
\subsection{Action-Angle Variables}

The Hamiltonian of the unperturbed problem, $\left(H_{01}\right)$ of Eq.~(\ref{eq:orderedhamiltonian}) is
\begin{equation}
H_{01} = -\frac{\Gamma_R\Gamma_r}{2\pi}\ln\left|\bfr\right|.
\end{equation}
Note that this depends only on the relative displacement of the dimer $\bfr$, so $x$ and $y$ are the only coordinates that vary along the unperturbed orbits.  Their equations of motion are
\begin{eqnarray}
\dot{x} &=& +\frac{1}{\Gamma_r}\frac{\partial H_0}{\partial y}
=-\frac{\Gamma_R}{2\pi}\frac{y}{x^2+y^2}\\
\dot{y} &=& -\frac{1}{\Gamma_r}\frac{\partial H_0}{\partial x}
=+\frac{\Gamma_R}{2\pi}\frac{x}{x^2+y^2}.
\end{eqnarray}
It immediately follows that
\begin{equation}
\frac{d}{dt}\left(x^2+y^2\right) = 2x\dot{x}+2y\dot{y} = 0,
\end{equation}
so that $|\bfr|^2=x^2+y^2$ is a constant of the unperturbed motion.

We define the action variable $J$ and the angle variable $\theta$ of the unperturbed motion,
\begin{equation}
J \equiv \frac{1}{2}\left|\bfr\right|^2 = \frac{1}{2}\left(x^2 + y^2\right), \; \; \;
\theta\equiv\arg\left(y + ix\right).
\end{equation}
We transform coordinates again, this time to eliminate $x$ and $y$ in favor of $\theta$ and $J$.  The Poisson bracket relations among the new coordinates may be calculated as follows,
\begin{equation}
\left\{\theta, J\right\} =
\frac{1}{\Gamma_r}\left(
\frac{\partial\theta}{\partial x}\frac{\partial J}{\partial y} -
\frac{\partial\theta}{\partial y}\frac{\partial J}{\partial x}
\right) =
\frac{1}{\Gamma_r}\left(
\frac{y}{x^2+y^2}\;y -
\frac{-x}{x^2+y^2}\;x
\right) =
\frac{1}{\Gamma_r}.
\end{equation}
It follows that $\theta$ and $J$ comprise a conjugate pair of coordinates, and that the Poisson tensor and its ordering remain unchanged if the coordinates are written,
\begin{equation}
\bfz = \langle\theta,X,x_j, \cdots, J, Y, y_j, \cdots\rangle.
\end{equation}

We next aim to write the Hamiltonian in the new coordinates.  Toward this end, we define the angles
\begin{equation}
\theta_j \equiv \arg\left[\left(Y-y_j\right) + i\left(X-x_j\right)\right],
\end{equation}
so that
\begin{equation}
\left(\bfR - \bfr_j\right)\cdot\bfr =
\sqrt{2J}\; \left|\bfR - \bfr_j\right|\; \cos\left(\theta - \theta_j\right).
\end{equation}
We also make use of the straightforwardly derived Fourier expansion
\begin{equation}
-\ln\left(1 \mp 2z\cos\theta+z^2\right) =
\sum_{\ell=1}^\infty\frac{2(\pm 1)^\ell}{\ell}\; \cos\left(\ell\theta\right) z^\ell.
\end{equation}
Then, after some algebra, the last two terms of Eq.~(\ref{eq:orderedhamiltonian}) transform to give us
\begin{eqnarray}
H
&=&
-\frac{\Gamma_R\Gamma_r}{4\pi}\ln\left(2J\right)
-\sum_{j\neq m,n}^N\sum_{k\neq j,m,n}^N\frac{\Gamma_j\Gamma_k}{4\pi}\ln \left|\bfr_j-\bfr_k\right|
-\sum_{j\neq m,n}^N\frac{\Gamma_R\Gamma_j}{2\pi}\ln\left|\bfR-\bfr_j\right|
\nonumber\\
& &
+
\sum_{\ell=2}^\infty\epsilon^\ell\;\frac{2}{\ell}
\left[
\frac{\Gamma_n^{\ell-1}+(-1)^\ell\Gamma_m^{\ell-1}}{\Gamma_R^{\ell-1}}
\right]
\sum_{j\neq m,n}^N\frac{\Gamma_r\Gamma_j}{4\pi}
\left(\frac{\sqrt{2J}}{\left|\bfR-\bfr_j\right|}\right)^\ell
\cos\left[\ell\left(\theta-\theta_j\right)\right].
\label{eq:finalhamiltonian}
\end{eqnarray}
We have thus succeeded in writing the Hamiltonian as
\begin{equation}
H = H_0 + \sum_{\ell=2}^\infty \epsilon^\ell H_\ell,
\end{equation}
where $H_0 = H_{01}+H_{02}$,
\begin{eqnarray}
H_{01} &=& -\frac{\Gamma_R\Gamma_r}{4\pi}\ln\left(2J\right),\label{eq:o0}\\
H_{02} &=&
-\sum_{j\neq m,n}^N\sum_{k\neq j,m,n}^N\frac{\Gamma_j\Gamma_k}{4\pi}\ln \left|\bfr_j-\bfr_k\right|
-\sum_{j\neq m,n}^N\frac{\Gamma_R\Gamma_j}{2\pi}\ln\left|\bfR-\bfr_j\right|,\label{eq:o1}\\
\noalign{\noindent \mbox{and for $\ell\geq 2$,}}\nonumber\\
H_\ell &=&
\frac{2}{\ell}
\left[
\frac{\Gamma_n^{\ell-1}+(-1)^\ell\Gamma_m^{\ell-1}}{\Gamma_R^{\ell-1}}
\right]
\sum_{j\neq m,n}^N\frac{\Gamma_r\Gamma_j}{4\pi}
\left(\frac{\sqrt{2J}}{\left|\bfR-\bfr_j\right|}\right)^\ell
\cos\left[\ell\left(\theta-\theta_j\right)\right].\label{eq:ol}
\end{eqnarray}
The rather simple form of the higher-order terms $H_\ell$ is encouraging.  To the extent that the unperturbed motion of $\bfr$ is oscillatory, these terms will be straightforward to average and integrate over unperturbed orbits.

It is very important to note that $\theta$ and $J$ are action-angle variables for the unperturbed problem with Hamiltonian $H_{01}$.  They are not action-angle variables for the full Hamiltonian $H$.  Our strategy now will be to use perturbation theory to find a near-identity canonical transformation so that $\theta$ and $J$ are action-angle variables to higher order in $\epsilon$.

\subsection{The Unperturbed Motion}

The Hamiltonian $H_{01}$ is independent of all coordinates other than $J$, and therefore the coordinates $\bfR$, $J$, and $\bfr_j$ for $j\neq m,n$ are constant along unperturbed orbits.  Only $\theta$ varies along unperturbed orbits according to
\begin{equation}
\dot{\theta} = \mathbb{P}_0dH_{01} = \mathbb{P}_0^{\theta J}\frac{\partial H_{01}}{\partial J} = -\frac{\Gamma_R}{4\pi J}.
\end{equation}
It follows that
\begin{equation}
\theta(t) = \theta_0-\Omega t,
\label{eq:unperturbed}
\end{equation}
where $\theta_0$ is a constant of integration, and where we have defined the rotation frequency
\begin{equation}
\Omega\equiv\frac{\Gamma_R}{4\pi J}.
\label{eq:RotationRate}
\end{equation}

This means that averages of a phase function $A$ along unperturbed orbits are simply averages over the angle variable $\theta$.  These are accomplished with the operator
\begin{equation}
\langle A\rangle\equiv\frac{1}{2\pi}\int_0^{2\pi}A(\theta)\; d\theta. \label{eq:avgphasefn}
\end{equation}
The oscillatory part of a phase function $A$ is then denoted
\begin{equation}
\widetilde{A}\equiv A-\left\langle A\right\rangle.
\end{equation}

\subsection{Averaging the Hamiltonian Over Unperturbed Orbits}

We are soon going to need the average of our Hamiltonian over unperturbed orbits, so we compute it here.  Because $H_{01}$ and $H_{02}$ are independent of $\theta$, they are unchanged by averaging over $\theta$; that is
\begin{eqnarray}
\langle H_{01}\rangle &=& H_{01},\\
\langle H_{02}\rangle &=& H_{02}.\\
\noalign{\noindent \mbox{
For $\ell\geq 2$, since $\left\langle\cos\left[\ell\left(\theta-\theta_j\right)\right]\right\rangle = 0$, it follows that
}}\nonumber\\
\langle H_\ell\rangle &=& 0. \label{eq:avHi}
\end{eqnarray}
The oscillatory parts of the Hamiltonian are then
\begin{eqnarray}
\widetilde{H}_{01} &=& 0\\
\widetilde{H}_{02} &=& 0\\
\widetilde{H}_\ell &=& H_\ell.
\end{eqnarray}

\section{Lie Transform Perturbation Theory}
\subsection{Overview}
At this point, our phase-space coordinates are $\bfz = \langle\theta,X,x_j, \cdots, J, Y, y_j, \cdots\rangle,$ where $j\neq m, n$.  We introduce the Lie derivative $\pounds_{\bfv}$, with respect to the vector field $\bfv$.  It acts on scalar fields $f\left(\bfz\right)$ and on other vector fields $\bfw\left(\bfz\right)$ in the following manner:
\begin{equation}
\pounds_{\bfv}f = v^if_{,i} = v^i\frac{\partial f}{\partial z^i},
\label{eq:LieDerivativeScalar}
\end{equation}
\begin{equation}
\left(\pounds_{\bfv}\bfw\right)^k = v^iw^k_{,i} - w^iv^k_{,i}.
\end{equation}

We can take the Lie derivative of any tensor by contracting this tensor with the requisite number of arbitrary 1-forms and vector fields, applying the Lie derivative to the resultant scalar, and then applying the Leibnitz rule.  We will need the Lie derivative of a Poisson tensor
\begin{equation}
\left(\pounds_{\bfv}\mathbb{P}\right)^{jk} = \mathbb{P}^{jk}_{,i}v^i - \mathbb{P}^{ik}v^j_{,i} - \mathbb{P}^{ji}v^k_{,i},
\label{eq:LieTensorDerivative}
\end{equation}
but the first term in Eq.~(\ref{eq:LieTensorDerivative}) is equal to zero, since our Poisson tensor has constant components.

The method of Lie transformations usually proceeds by picking a generating function, g, which gives a vector field
\begin{equation}
v^j = \mathbb{P}^{jk}g_{,k}  \label{eq:scalargenerator}
\end{equation}
and noting that the following coordinate transformation is always canonical
\begin{equation}
\mybar{z} = \exp\left(+\pounds_{\bfv}\right) z.
\end{equation}
That is, the Lie derivative of $\mathbb{P}$ disappears under such a generating function, thereby preserving the symplectic structure.  While this choice, Eq.~(\ref{eq:scalargenerator}),  overdetermines the generating vector field, we will use it until a little more subtlety is need to address fourth-order corrections.  This transformation results in a new Hamiltonian
\begin{equation}
\mybar{H} = \exp\left(-\pounds_{\bfv}\right) H.
\end{equation}
If $\bfv$ is of order $\epsilon^j$, the transformation will affect the Hamiltonian only at order $\epsilon^j$ and higher.  

Our strategy will be to determine successive vector generators $\bfv_i$, starting at second order, that preserve the Poisson tensor and eliminate the Hamiltonian at all higher orders.  To leave the largest variety of near identity transformations at our disposal, we choose an infinite product form, in the manner of Dragt and Finn~\cite{bib:DragtFinn}:  
\begin{equation}
\mybar{\bfz} = 
\exp\left(+\epsilon^2\pounds_2\right)\exp\left(+\epsilon^3\pounds_3\right)\exp\left(+\epsilon^4\pounds_4\right)\cdots
\bfz.
\label{eq:oalltranform}
\end{equation}
Here, $\pounds_n \equiv \pounds_{\bfv_n}$.  This results in a new Hamiltonian and Poisson tensor:
\begin{eqnarray}
\mybar{H} &=&
\cdots\exp\left(-\epsilon^4\pounds_4\right)\exp\left(-\epsilon^3\pounds_3\right)\exp\left(-\epsilon^2\pounds_2\right)
H, \label{eq:HamTrans} \\
\mybar{\mathbb{P}} &=&
\cdots\exp\left(-\epsilon^4\pounds_4\right)\exp\left(-\epsilon^3\pounds_3\right)\exp\left(-\epsilon^2\pounds_2\right)
\mathbb{P}. \label{eq:PoisTrans}
\end{eqnarray}

\subsection{Ordering}

\subsubsection{Poisson Tensor Ordering Redux}

Now that we have the Lie derivative machinery at our disposal, we can justify our use~Eq.(\ref{eq:PoissonOrdering}) for the initial Poisson ordering.  Consider the Lie derivative of a scalar $f$,~Eq.(\ref{eq:LieDerivativeScalar}), with respect to a vector field,~Eq.(\ref{eq:scalargenerator}), generated by the scalar $g$: 
\begin{equation}
\pounds_{\bfv}f = \mathbb{P}^{jk}\frac{\partial g}{\partial z^k}\frac{\partial f}{\partial z^j}.
\end{equation}
Because of the symplectic structure of the Poisson tensor, we can partition the terms on the right hand side into two groups: those that have derivatives with respect to variables $\langle\theta,J\rangle$, and those that have derivatives with respect to variables $\langle X,x_j, \cdots, Y, y_j, \cdots\rangle$.  For the Hamiltonians and scalar generators we are considering, the magnitude of the $\langle\theta,J\rangle$ terms in the Lie derivative will be reduced by roughly a factor of $J$ when compared to the magnitude of $g$ and $f$ together.  Similarly, the magnitude of the $\langle X,x_j, \cdots, Y, y_j, \cdots\rangle$ terms in the Lie derivative will be reduced by roughly a factor of $\left|\bfR-\bfr_j\right|^2$ or $\left|\bfr_i-\bfr_j\right|^2$.  Therefore the ratio of the $\langle X,x_j, \cdots, Y, y_j, \cdots\rangle$ terms in the Lie derivative to the  $\langle\theta,J\rangle$ terms will be of order $\epsilon^2$.  This relative ordering can be formally achieved by ordering the initial Poisson tensor in the manner given by~Eq.(\ref{eq:PoissonOrdering}):
\begin{equation}
\pounds_{\bfv}f = \left(\mathbb{P}_0^{jk} + \epsilon^2 \mathbb{P}_2^{jk}\right)\frac{\partial g}{\partial z^k}\frac{\partial f}{\partial z^j} = \mathbb{P}_0^{jk}\frac{\partial g}{\partial z^k}\frac{\partial f}{\partial z^j} + \epsilon^2 \mathbb{P}_2^{jk}\frac{\partial g}{\partial z^k}\frac{\partial f}{\partial z^j}.
\end{equation}

\subsubsection{Lie Transform Ordering}

Expanding each of the exponentials of~Eq.(\ref{eq:HamTrans} \& \ref{eq:PoisTrans}) to fourth order and putting in the explicit initial ordering of the Hamiltonian and Poisson tensor gives:
\begin{eqnarray}
\mybar{H} &=& \left(1-\epsilon^4\pounds_4 \right)\left(1-\epsilon^3\pounds_3 \right)\left(1-\epsilon^2\pounds_2 + \epsilon^4\frac{1}{2}\pounds_2^2\right)\left(H_0 + \sum_{n=2}\epsilon^n H_n\right) \label{eq:hexpansion}\\
\mybar{\mathbb{P}} &=& \left(1-\epsilon^4\pounds_4 \right)\left(1-\epsilon^3\pounds_3 \right)\left(1-\epsilon^2\pounds_2 + \epsilon^4\frac{1}{2}\pounds_2^2\right)\left(\mathbb{P}_0 + \epsilon^2 \mathbb{P}_2\right) \label{eq:hbarexpansion}
\end{eqnarray}
We suppose that both the final Hamiltonian and Poisson tensor are also ordered in $\epsilon$.  Then, expansion of the above in powers of $\epsilon$ and matching terms yields the sequence of equations
\begin{eqnarray}
\mybar{H}_0 &=& H_0\label{eq:order0}\\
\mybar{H}_1 &=& 0\label{eq:order1}\\
\mybar{H}_2 &=& H_2 - \pounds_2 H_0\label{eq:order2}\\
\mybar{H}_3 &=& H_3 - \pounds_3 H_0\label{eq:order3}\\
\mybar{H}_4 &=& H_4 - \pounds_4 H_0  - \pounds_2 H_2 + \frac{1}{2} \pounds^2_2 H_0  \label{eq:order4}\\
&\vdots&\nonumber
\end{eqnarray}
and
\begin{eqnarray}
\mybar{\mathbb{P}}_0 &=& \mathbb{P}_0\label{eq:porder0}\\
\mybar{\mathbb{P}}_1 &=& 0\label{eq:porder1}\\
\mybar{\mathbb{P}}_2 &=& \mathbb{P}_2 - \pounds_2 \mathbb{P}_0\label{eq:porder2}\\
\mybar{\mathbb{P}}_3 &=&  - \pounds_3 \mathbb{P}_0\label{eq:porder3}\\
\mybar{\mathbb{P}}_4 &=& - \pounds_4 \mathbb{P}_0  - \pounds_2 \mathbb{P}_2 + \frac{1}{2} \pounds^2_2 \mathbb{P}_0  \label{eq:porder4}\\
&\vdots&\nonumber
\end{eqnarray}

After applying the Lie transforms, we would like the equations of motion to consist of the motion of the dimer and the motion of all the other vortices with the dimer.  That is, $ \dot{\mybar{\bfz}} = \mathbb{P}_0 dH_0+\mathbb{P}_2 dH_0 $.  We can ensure this by requiring that order by order, starting at second order, the transformed Poisson tensor remains unchanged and the transformed Hamiltonian disappears.  Furthermore, secular perturbation theory forbids us from including any averaged, Eq.~(\ref{eq:avgphasefn}), contributions to our transformed Hamiltonians in the calculation of the vector generators.  This means that we will need to check that  $\langle \mybar{H}_\ell\rangle = 0$ at each order.  These considerations lead to the following conditions:
\begin{eqnarray}
\widetilde{H_2} - \widetilde{\pounds_2 H_0}   &=& 0 \label{eq:CondOrder2}\\
\widetilde{H_3} - \widetilde{\pounds_3 H_0}   &=& 0 \label{eq:CondOrder3}\\
\widetilde{H_4} - \widetilde{\pounds_4 H_0}  - \widetilde{\pounds_2 H_2} + \frac{1}{2} \widetilde{\pounds^2_2 H_0} &=&  0 \label{eq:CondOrder4}
\end{eqnarray}
and
\begin{eqnarray}
\pounds_2 \mathbb{P}_0 &=& 0 \label{eq:PCondOrder2}\\
\pounds_3 \mathbb{P}_0 &=& 0 \label{eq:PCondOrder3}\\
\pounds_4 \mathbb{P}_0  + \pounds_2 \mathbb{P}_2 - \frac{1}{2} \pounds^2_2 \mathbb{P}_0  &=& 0. \label{eq:PCondOrder4}
\end{eqnarray}

\subsubsection{Orders Two and Three}

To maintain the second-order motion that results from the interaction of the non-dimer vortices, as well as their interaction with the dimer as though it were a single particle of circulation $\Gamma_R$, we must demand that the conditions in Eq.~(\ref{eq:CondOrder2}) \& (\ref{eq:PCondOrder2}) are met.  The second of these conditions is trivially fulfilled if
\begin{equation}
v_2^i = \mathbb{P}_0^{ik}g_{2,k}.
\end{equation}
Thus the vector generator at second order, which is now completely determined by a scalar generator $g_2$, has only two non-zero components, 
\begin{equation}
\bfv_2 = \langle \frac{1}{\Gamma_r}\frac{\partial g_2}{\partial J}, 0, \cdots, -\frac{1}{\Gamma_r}\frac{\partial g_2}{\partial \theta}, 0, \cdots \rangle.  
\end{equation}
This allows us to write the condition, Eq.~(\ref{eq:CondOrder2}), on our Hamiltonian as:
\begin{equation}
\widetilde{H_2}-\frac{\Gamma_R}{4\pi J}\widetilde{\frac{\partial g_2}{\partial\theta}} = 0. \label{eq:H2trans}
\end{equation}
If we demand that $g_2$ be single-valued to preclude secular terms and consider Eq.~(\ref{eq:avHi}), it is immediately apparent that $\langle \mybar{H}_2\rangle = \langle H_2\rangle - \langle \frac{\Gamma_R}{4\pi J}\frac{\partial g_2}{\partial\theta} \rangle = 0$.  Now we must solve Eq.~(\ref{eq:H2trans}) for the generator
\begin{equation}
g_2 = \frac{4\pi J}{\Gamma_R} \int^{\theta} \widetilde{H}_2\left(\theta'\right)\; d\theta'.
\end{equation}
From Eq.~(\ref{eq:ol}), this yields the generator
\begin{equation}
g_2 = \frac{\Gamma_r}{\Gamma_R}\;J^2
\sum_{j\neq m,n}^N\Gamma_j\;
\frac{\sin\left(2\theta-2\theta_j\right)}{\left|\bfR-\bfr_j\right|^2}.
\end{equation}

At this point it would be ideal to be able to renormalize the unperturbed orbits by treating all of $H_0$ as if it were the new unperturbed problem, in the manner of a superconvergent Lie transformation.  However, this problem is already non-integrable and therefore not amenable to this treatment.  So, we proceed to third order with the same unperturbed motion. 
Identical considerations lead to the following third-order generator
\begin{equation}
g_3 = \left(\frac{2}{3}\right)^2\frac{\sqrt{2}\Gamma_r\left(\Gamma_n-\Gamma_m\right)}{\Gamma_R^2}\;J^{\frac{5}{2}}
\sum_{j\neq m,n}^N\Gamma_j\;
\frac{\sin\left(3\theta-3\theta_j\right)}{\left|\bfR-\bfr_j\right|^3}.
\end{equation}

The generator $g_2$ provides the most important correction term present.  With it, the transformation given by Eq.~(\ref{eq:oalltranform}) takes us to a set of coordinates in which the dimer may be approximated as a single vortex of circulation $\Gamma_R$ and position $\bfR$.  Whereas this approximation was valid only to order $\epsilon$ in our original coordinates, in the new coordinates it is accurate to order $\epsilon^2$ -- that is, the first corrections to it are order $\epsilon^3$.  When the generator $g_3$ is included the validity of this approximation is extended to order $\epsilon^3$. We next refine this transformation to push the corrections to still higher order.

\subsubsection{Order Four}

Using $\pounds_2\mathbb{P}_0 = 0$ and $H_2-\pounds_2H_0 = 0$, the conditions at fourth order, Eq.~(\ref{eq:CondOrder4}) \& (\ref{eq:PCondOrder4}), can be written
\begin{equation}
\widetilde{H_4} - \widetilde{\pounds_4 H_0}  - \frac{1}{2}\widetilde{\pounds_2 H_2} =  0 \label{eq:smallCondOrder4}
\end{equation}
and
\begin{equation}
\pounds_4\mathbb{P}_0 + \pounds_2\mathbb{P}_2 = 0.
\end{equation}
This latter equation is satisfied with the choice
\begin{equation}
v_4^j = \mathbb{P}_0^{jk}g_{4,k} + \mathbb{P}_2^{jk}g_{2,k}.
\end{equation}
Our vector generator at fourth order now has the form
\begin{equation}
\bfv_4 = \langle \frac{1}{\Gamma_r}\frac{\partial g_4}{\partial J}, \frac{1}{\Gamma_R}\frac{\partial g_2}{\partial Y}, \frac{1}{\Gamma_j}\frac{\partial g_2}{\partial y_j}, \cdots, 
-\frac{1}{\Gamma_r}\frac{\partial g_4}{\partial \theta}, -\frac{1}{\Gamma_R}\frac{\partial g_2}{\partial X}, -\frac{1}{\Gamma_j}\frac{\partial g_2}{\partial x_j}, \cdots \rangle.  
\end{equation}
All but two of these terms are already determined from $g_2$.  We will fix $g_4$ by considering Eq.~(\ref{eq:smallCondOrder4}).  Once again we demand that $g_4$ is single-valued and check for any averaged contributions to $\mybar{H}_4$.  Unlike at second and third order, we encounter an averaged contribution to $\mybar{H}_4$, due to $\pounds_2 H_2$, that secular perturbation theory forbids us from transforming away.  Before we consider how this averaged term affects our perturbation theory, we can derive $g_4$ from the purely oscillatory parts of each term:
\begin{equation}
g_4 = \frac{4\pi J}{\Gamma_R} \int^{\theta} \left(\widetilde{H}_4  - \widetilde{\mathbb{P}_2^{jk}g_{2,k}H_{0,j}} - \frac{1}{2}\widetilde{\mathbb{P}_0^{jk}g_{2,k}H_{2,j}}\right)\; d\theta'.
\end{equation}
From this the fourth-order generator is straightforwardly, though laboriously, derived:
\begin{equation}
\begin{split}
g_4 \; = \;
& \frac{\Gamma_r}{4\Gamma_R^2}\;J^3
\sum_{j\neq m,n}^N\Gamma_j\; 
\left[
\left(2\left( \frac{\Gamma_n^3+\Gamma_m^3}{\Gamma_R^2} \right)-\Gamma_j \right) \frac{\sin\left(4\theta-4\theta_j\right)}{\left|\bfR-\bfr_j\right|^4} 
\; + \; 8\left(\Gamma_R+\Gamma_j\right) \frac{\sin\left(2\theta-2\theta_j\right)}{\left|\bfR-\bfr_j\right|^4}
\bracketnewln{+}\sum_{k\neq j,m,n}^N\Gamma_k\; \left(
8\frac{\sin\left(2\theta-3\theta_j+\theta_k\right)}{\left|\bfR-\bfr_j\right|^3\left|\bfR-\bfr_k\right|} 
\;-\; 8\frac{\sin\left(2\theta-3\theta_j+\theta_{jk}\right)}{\left|\bfR-\bfr_j\right|^3\left|\bfr_j-\bfr_k\right|} 
-\frac{\sin\left(4\theta-2\theta_j-2\theta_k\right)}{\left|\bfR-\bfr_j\right|^2\left|\bfR-\bfr_k\right|^2} \right) \right],
\end{split}
\end{equation}
where $\theta_{jk} \equiv \arg\left[\left(y_j-y_k\right) + i\left(x_j-x_k\right)\right]$.

\subsubsection{Averaged Contribution to $\mybar{H}_4$}

At fourth order, we must consider the effect of the non-oscillatory term due to $\pounds_2 H_2$.  This gives us an averaged contribution to the Hamilton, 
\begin{equation}
\left\langle \mybar{H}_4\right\rangle = \frac{3\Gamma_r J^2}{4 \pi \Gamma_R}\;
\sum_{j\neq m,n}^N \sum_{k\neq m,n}^N\Gamma_j\Gamma_k\;
\frac{\cos\left(2\theta_j-2\theta_k\right)}{\left|\bfR-\bfr_j\right|^2\left|\bfR-\bfr_k\right|^2},
\label{eq:h4contrib}
\end{equation}
which changes the dynamics
\begin{equation}
\dot{\mybar{z}^j} = \mathbb{P}_0^{jk} \frac{\partial H_{01}}{\partial z^k} + \epsilon^2 \mathbb{P}_2^{jk} \frac{\partial H_{02}}{\partial z^k} + \epsilon^4\mathbb{P}_0^{jk} \frac{\partial \left\langle \mybar{H}_4\right\rangle}{\partial z^k}. \label{eq:FinalDynamics}
\end{equation}
Only the angle variable of the dimer, $\theta$, is affected at fourth order.  In particular, $J$ remains a constant of the motion and 
 \begin{equation}
\dot{\theta} = -\Omega + \epsilon^4\frac{3J}{2 \pi \Gamma_R}\;
\sum_{j\neq m,n}^N \sum_{k\neq m,n}^N\Gamma_j\Gamma_k\;
\frac{\cos\left(2\theta_j-2\theta_k\right)}{\left|\bfR-\bfr_j\right|^2\left|\bfR-\bfr_k\right|^2}. \label{eq:freqcorrection}
\end{equation}

	Thus the rotational frequency of the dimer decreases at fourth order.  Though the extent to which it does decrease depends on the configuration of other vortices.  To help interpret Eq.~(\ref{eq:freqcorrection}), we can write the unit vector from vortex $k$ to the dimer as $\hat{\bfR}_k = \frac{\bfR - \bfr_k}{\left|\bfR-\bfr_k\right|}$ and use the $\hodge{}$ operator from Eq.~(\ref{eq:InitialDynamics}), equivalent to rotating the vector by $\frac{\pi}{2}$, which gives the following for the important part of the frequency correction:
 \begin{equation}
\frac{\cos\left(2\theta_j-2\theta_k\right)}{\left|\bfR-\bfr_j\right|^2\left|\bfR-\bfr_k\right|^2} = 
\frac{\left[ \hat{\bfR}_j \cdot \left(\hat{\bfR}_k+\hodge{\hat{\bfR}_k} \right)\right]\left[ \hat{\bfR}_k \cdot \left(\hat{\bfR}_j+\hodge{\hat{\bfR}_j} \right)\right] }{\left|\bfR-\bfr_j\right|^2\left|\bfR-\bfr_k\right|^2}. \label{eq:freqcorrection2}
\end{equation}

	Though it is not obvious, any configuration of any number of extra vortices will always decrease or maintain the dimer's rotational frequency.  For a single extra vortex, with either positive or negative circulation, this is apparent.  The complications arise when considering the three-body contributions to~Eq.(\ref{eq:freqcorrection}).  We can see, from~Eq.(\ref{eq:freqcorrection2}), that two extra, like-signed vortices which are co-linear with the dimer will further decrease the rotation rate.  If, however, they are both an equal distance from the dimer and are at right angles to each other with respect to the dimer, there will be zero net change in the dimer's rotation rate.  The results of these two scenarios are interchanged when the two extra vortices have opposite signed circulations.  When there are more than two extra vortices, it becomes more difficult to argue that the change in the dimer's rotation rate is negative definite, though we have not seen a counter example in numerical simulations. \\

	As a physics analogy, we consider $\theta$ to be an internal variable of the dimer, much like spin.  Up until now the rest of the system has been transparent to the spin of the dimer.  With the addition of the fourth-order averaged Hamiltonian, we can view the spin as coupling to a new field; a field generated by the vortices, but coupled to only by vortices with non-zero spin, i.e., the vortex dimers.  This new field gives the dimer a position-dependent effective spin, which is smaller than the bare spin ($\Omega$).  This interesting view remains valid through at least fifth order in our calculations.	\\

\subsection{Transformed Variables}

To fourth order, the transformation given in Eq.~{\ref{eq:oalltranform}} is written as
\begin{equation}
{\mybar{\bfz} \brace \bfz} = 
\left(1\pm \epsilon^2 \pounds_2 \pm \epsilon^3 \pounds_3  + \epsilon^4 \left(\pm\pounds_4+\frac{1}{2}\pounds_2^2 \right)\right)
{\bfz \brace \mybar{\bfz}},
\end{equation}
and results in the following specific transformations, where $\alpha_j \equiv \left(2\frac{\Gamma_n^3+\Gamma_m^3}{\Gamma_R^2} - \Gamma_j \right)$ and $\beta_j \equiv  \left(\Gamma_R+\Gamma_j\right)$.  To get the forward transformation, simply use the upper operator of $\pm$ or $\mp$; to get the backward transformation, use the lower of $\pm$ or $\mp$ and replace all transformed coordinates $\mybar{\bfz}$ with the original coordinates $\bfz$ and vice-versa.

\begin{equation}
\begin{split}
\mybar{\theta} = \; \theta \; & \pm \; \epsilon^2\frac{2}{\Gamma_R}J \sum_{j\neq m,n}^N\Gamma_j   \frac{\sin\left(2\theta-2\theta_j\right)}{\left|\bfR-\bfr_j\right|^2} \\
& \; \pm \; \epsilon^3\frac{10 \sqrt{2}}{9}\frac{\left(\Gamma_n-\Gamma_m\right)}{\Gamma_R^2} J^{\frac{3}{2}} \sum_{j\neq m,n}^N\Gamma_j\frac{\sin\left(3\theta-3\theta_j\right)}{\left|\bfR-\bfr_j\right|^3} \\
& \; \pm \; \epsilon^4\frac{1}{\Gamma_R^2}J^2 \sum_{j\neq m,n}^N\Gamma_j\; 
\left[  \left(\pm\Gamma_j+\frac{3}{4}\alpha_j \right)\frac{\sin\left(4\theta-4\theta_j\right)}{\left|\bfR-\bfr_j\right|^4} 
\;+\;6\beta_j \frac{\sin\left(2\theta-2\theta_j\right)}{\left|\bfR-\bfr_j\right|^4} \right. \\
& \left. \; \; \; \; \; \; \; \;\; \; \; \; \; \; \; \;\; \; \; \; \; \; \; \; \;+\;\sum_{k\neq j,m,n}^N\Gamma_k \left(
6\frac{\sin\left(2\theta-3\theta_j+\theta_k\right)}{\left|\bfR-\bfr_j\right|^3\left|\bfR-\bfr_k\right|}
\;-\; 6\frac{\sin\left(2\theta-3\theta_j+\theta_{jk}\right)}{\left|\bfR-\bfr_j\right|^3\left|\bfr_j-\bfr_k\right|} \right. \right. \\
& \left. \left. \; \; \; \; \; \; \; \;\; \; \; \; \; \; \; \; \; \; \; \; \; \; \; \;\; \; \; \; \; \; \; \;\; \; \; \; \; \; \; \;\; \; \; \; \; \; \; \; \;+\;\left(\pm1-\frac{3}{4}\right)\frac{\sin\left(4\theta-2\theta_j-2\theta_k\right)}{\left|\bfR-\bfr_j\right|^2\left|\bfR-\bfr_k\right|^2} \right) \right]	\label{eq:Thetatransform}
\end{split}
\end{equation}

\begin{equation}
\begin{split}
\mybar{J} = \; J \; & \:\mp \; \epsilon^2\frac{2}{\Gamma_R}J^2 \sum_{j\neq m,n}^N\Gamma_j   \frac{\cos\left(2\theta-2\theta_j\right)}{\left|\bfR-\bfr_j\right|^2} \\
& \;\mp\; \epsilon^3\frac{4 \sqrt{2}}{3}\frac{\left(\Gamma_n-\Gamma_m\right)}{\Gamma_R^2} J^{\frac{5}{2}} \sum_{j\neq m,n}^N\Gamma_j\frac{\cos\left(3\theta-3\theta_j\right)}{\left|\bfR-\bfr_j\right|^3} \\
& \;\mp\; \epsilon^4\frac{1}{\Gamma_R^2}J^3 \sum_{j\neq m,n}^N\Gamma_j\; 
\left[  \alpha_j\frac{\cos\left(4\theta-4\theta_j\right)}{\left|\bfR-\bfr_j\right|^4} 
\;+\;4\beta_j \frac{\cos\left(2\theta-2\theta_j\right)}{\left|\bfR-\bfr_j\right|^4}
\;\mp \frac{4 \Gamma_j}{\left|\bfR-\bfr_j\right|^4} \right. \\
& \left. \; \; \; \; \; \; \; \; \;\; \; \; \; \; \; \; \;\; \; \; \; \; \; \; \;+\;\sum_{k\neq j,m,n}^N\Gamma_k \left(
4\frac{\cos\left(2\theta-3\theta_j+\theta_k\right)}{\left|\bfR-\bfr_j\right|^3\left|\bfR-\bfr_k\right|} 
\;-\;4\frac{\cos\left(2\theta-3\theta_j+\theta_{jk}\right)}{\left|\bfR-\bfr_j\right|^3\left|\bfr_j-\bfr_k\right|} \right. \right. \\
& \left. \left.\; \; \; \; \; \; \; \; \; \; \; \; \; \; \; \; \; \; \; \;\; \; \; \; \; \; \; \;\; \; \; \; \; \; \; \;\; \; \; \; \; \; \; \; \; \; \; \; \mp4\frac{\cos\left(2\theta_j-2\theta_k\right)}{\left|\bfR-\bfr_j\right|^2\left|\bfR-\bfr_k\right|^2}
\;-\; \frac{\cos\left(4\theta-2\theta_j-2\theta_k\right)}{\left|\bfR-\bfr_j\right|^2\left|\bfR-\bfr_k\right|^2} \right) \right]	\label{eq:Jtransform}
\end{split}	
\end{equation}

\begin{equation}
\begin{split}
\mybar{X} =  \; X \; & \mp \; \epsilon^4\frac{2\Gamma_r}{\Gamma_R^2}J^2 \sum_{j\neq m,n}^N\Gamma_j\;  \frac{\sin\left(2\theta-3\theta_j\right)}{\left|\bfR-\bfr_j\right|^3} \\
\mybar{Y} =  \; Y\; & \pm \; \epsilon^4\frac{2\Gamma_r}{\Gamma_R^2}J^2 \sum_{j\neq m,n}^N\Gamma_j\;   \frac{\cos\left(2\theta-3\theta_j\right)}{\left|\bfR-\bfr_j\right|^3}\\
\mybar{x}_i =  \;  x_i \;&  \pm\; \epsilon^4\frac{2\Gamma_r}{\Gamma_R}J^2  \frac{\sin\left(2\theta-3\theta_i\right)}{\left|\bfR-\bfr_i\right|^3} \\
 \mybar{y}_i = \;  y_i\;& \mp \; \epsilon^4\frac{2\Gamma_r}{\Gamma_R}J^2  \frac{\cos\left(2\theta-3\theta_i\right)}{\left|\bfR-\bfr_i\right|^3}	\label{eq:XYtransform}
\end{split}
\end{equation}

\section{Numerical Results}
We propose to use the above-described transformation as the basis of a numerical method, whereby we make this transformation, integrate the reduced system, and invert the transformation.  The overall efficacy of this Lie transform method will depend to a large degree on various factors such as the number, configuration, and circulation of vortices in the system.  To probe the essential behavior of this method, we will consider the simplest case, that of three vortices of equal circulation, two of which are much closer to one another.  This enables us to unambiguously state that our small perturbation parameter is  $\epsilon = |\bfr|/|\bfR-\bfr_j|$.  For reference, we used a Runge-Kutta-Fehlberg (4, 5) adaptive time-stepping numerical integrator.

\subsection{Efficiency}

\begin{figure}[htbp]
  \begin{center}
    \input{Efficiency.tex}
    \caption{The vertical axis shows the ratio of the number of calls to the adaptive time-stepper for the Regular Method $\left(N_R\right)$ to the Lie Method $\left(N_L\right)$, i.e. the speed-up factor.  This is plotted against  $\epsilon,$ the vortex dimer separation $\left|\bfr_n-\bfr_m \right|$ divided by the separation between the remaining vortex and the dimer $\left|\bfR-\bfr_j\right|.$  \; A log-log version of this graph is inset.  This corresponds to a power law $\frac{N_R}{N_L} \propto \epsilon^{\alpha}$ where $\alpha \approx -1.8$}
    \label{graph:graph1}
  \end{center}
\end{figure}
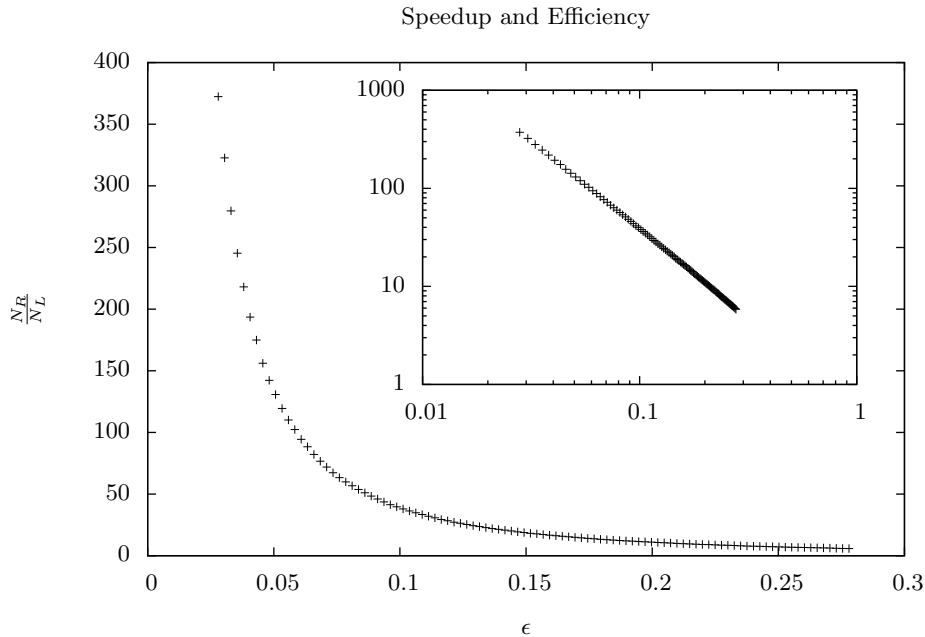

We would like to measure how much faster an algorithm based on the dynamics Eq.~(\ref{eq:FinalDynamics}) and transformations Eq.~(\ref{eq:Thetatransform} - \ref{eq:XYtransform}) would run when compared to one based on the original point-vortex dynamics Eq.~(\ref{eq:InitialDynamics}).  The pertinent quantities we want to track are the number of calls each method makes to the base adaptive timer stepper.  In particular, we will consider the ratio of these two numbers, i.e., the speedup factor, and how it behaves when $\epsilon$ becomes small and our transformations therefore become more applicable.  Since the time that it takes a vortex in the dimer to rotate through a given angle scales as the dimer separation squared Eq.~(\ref{eq:unperturbed}, \ref{eq:RotationRate}), we would roughly expect the speedup factor to scale as $\epsilon^{-2}$ for all orders of Lie transforms.  As one can see in Figure \ref{graph:graph1}, this is very nearly the case.  The magnitude of the speedup factor will certainly depend on the tolerance that one sets.  However, increasing or decreasing the tolerance will not affect how the speedup factor scales with $\epsilon$.  Neither will this scaling be modified by the contributions to the Lie transform algorithm run-time from the transformations themselves.  While these do involve pairwise sums, they are calculated only when saving the state of the system is needed, and therefore don't constitute anything worse than a small multiplicative contribution to the overall run-time.  Indeed their inclusion is only an additive contribution to the run-time if you are interested only in initial and final vortex configurations.

\subsection{Accuracy}

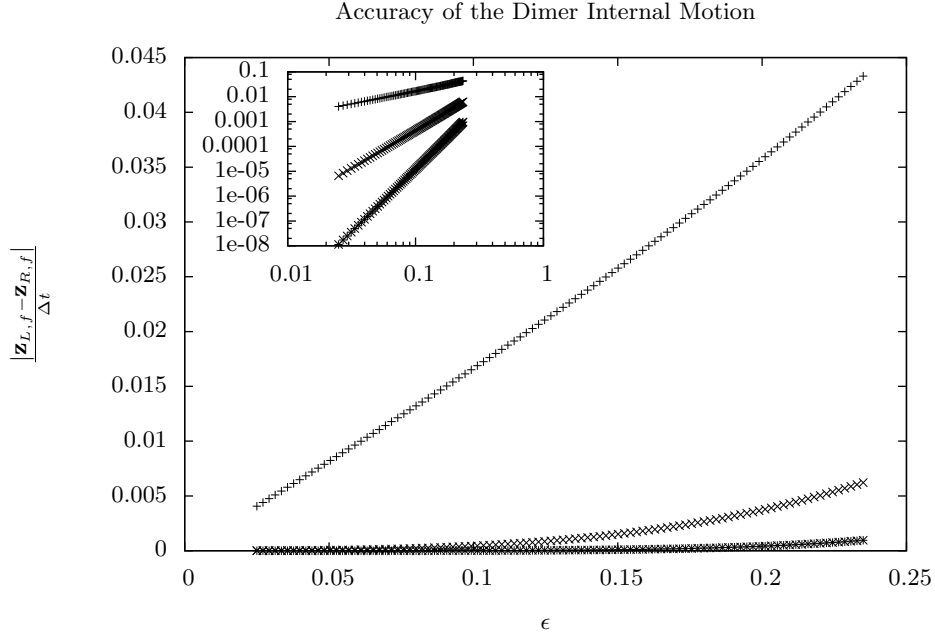
\begin{figure}[htbp]
  \begin{center}
    \input{AccuracyD.tex}
    \caption{The vertical axis shows the natural norm of the difference between the final configuration space points evolved under the Lie and Regular methods, normalized by the time elapsed.  In particular, in this graph we consider the positions of only the dimer vortices with respect to their center of circulation.  The three lines, from top to bottom, represent the Lie transform at 0th, 2nd, and 4th orders respectively.  The inset log-log plot shows that they all follow power laws with exponents 1, 3, and 5 respectively.}
    \label{graph:graph2}
  \end{center}
\end{figure}

\begin{figure}[htbp]
  \begin{center}
    \input{AccuracyR.tex}
    \caption{The vertical axis shows the natural norm of the difference between the final configuration space points evolved under the Lie and Regular methods, normalized by the time elapsed.  In particular, in this graph we consider the positions of only the dimer and other vortex.  The two lines, from top to bottom, represent the Lie transform at 0th - 2nd and 4th orders respectively.  The inset log-log plot shows that they both follow power laws with exponents 2 and 4 respectively.}
    \label{graph:graph3}
  \end{center}
\end{figure}
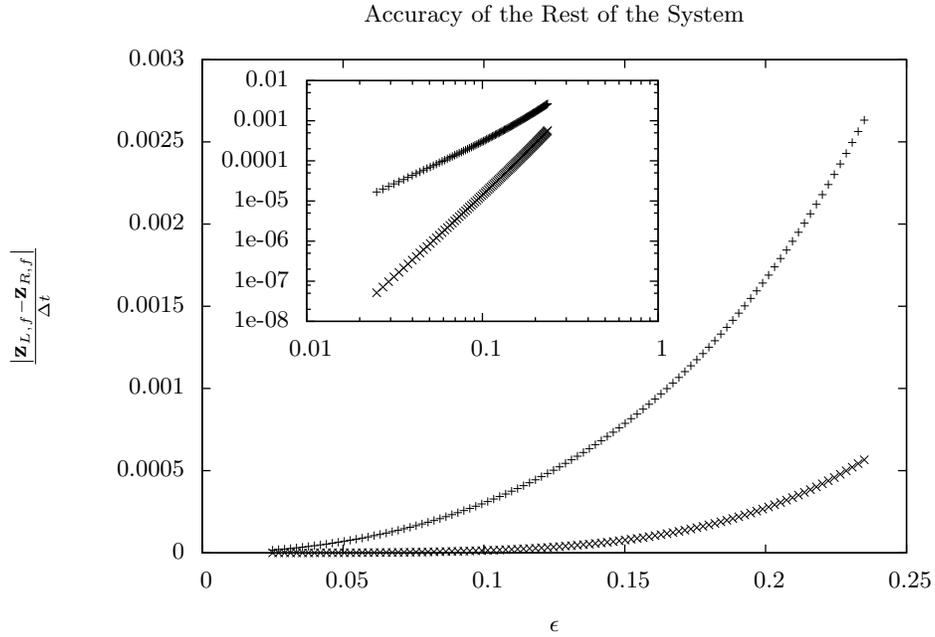

We characterize the accuracy of the Lie transform algorithm by comparing how two identical point-vortex systems differ after evolving one with the regular method, and the other with the Lie method.  Their deviation is quantified by the natural norm of the difference of their positions in configuration space.  To have the regular method be a stand-in for the actual motion we must set the tolerance to be very small, which forces us to consider only the deviations that occur at small times.  These deviations scale linearly with time, so the quantity of interest is really the deviation divided by time.  Furthermore, there is a natural separation of the overall time-normalized deviation into that of the positions of the dimer vortices with respect to their center of circulation and that of the positions of the dimer itself and extra vortex.  The time-normalized deviation of the first group is plotted against $\epsilon$ in figure \ref{graph:graph2} for three orders of the Lie transformation, while that of the second group is plotted in a similar manner in figure \ref{graph:graph3}, though for only two orders of the Lie transformation.  Notice that because the constituents of the vortex pair have equal circulation, there is no 3rd order transformation to consider.  The most important thing to note about each graph is that the time-normalized deviation scales as $\epsilon^{\alpha}$, where $\alpha$ increases with the order of Lie transform used.  That is, the Lie method gets more accurate as the dimer pair separation decreases and as we include higher orders of the Lie transform.

\subsection{Energy Conservation}

\begin{figure}[htbp]
  \begin{center}
    \input{Energy.tex}
    \caption{The vertical axis of this log-log plot shows two different items.  The first is the change in energy per unit time, $\frac{dH}{dt}$, for evolution with the regular method.  This corresponds to the line which increases as $\epsilon$ decreases. Note that this is a power law, $\frac{dH}{dt} \propto \epsilon^{\alpha}$, with $\alpha$ of roughly -3.  The second item is the amplitude, $A$, of oscillations in the energy about the fixed average for evolution with the Lie method.  This corresponds to the three line which decrease as $\epsilon$ decreases.  Note that each follows a power law $A \propto \epsilon^{\alpha}$.  From top to bottom, which corresponds to the Lie method at 0th, 2nd, and 4th orders, $\alpha$ is roughly 2, 4, and 6 respectively.}
    \label{graph:graph4}
  \end{center}
\end{figure}
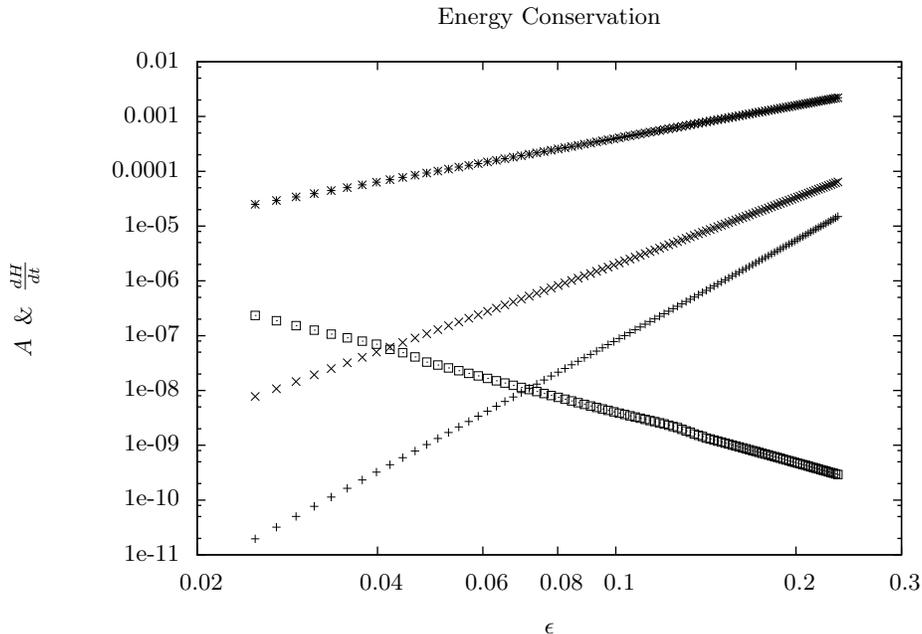

With respect to the close vortices circulating in a dimer, most numerical integration methods will systematically accrue errors that increase their separation.  This is equivalent to decreasing the total energy, which is forbidden in an autonomous Hamiltonian system Eq.~(\ref{eq:EnergyConservation}).  With the regular integration method, the decrease in total energy per unit time is roughly proportional to the specified tolerance.  More importantly, we can see from Figure \ref{graph:graph4} that this change in total energy per unit time roughly sales as $\epsilon^{-3}$.  This indicates that the non-constancy of the energy will certainly become a problem at small separations, no mater the tolerance.  The performance of our Lie method is in marked contrast to that of the regular method.  At every order of the Lie transform, the average energy remains constant.  On top of this desired constant energy there are sinusoidal oscillations with a frequency commensurate with that of the dimer's rotation and an amplitude that depends on the order of the Lie transform and $\epsilon$.  As indicated in Figure \ref{graph:graph4} the amplitude scales as $A \propto \epsilon^{\alpha}$.  For the Lie transform at orders 0, 2, and 4, we have that $\alpha$ is nearly 2, 4, and 6 respectively.  Thus the accuracy of energy conservation under each order of the Lie transform method becomes better as $\epsilon$ decreases.

\subsection{Algorithmic Concerns}	

In our implementation of this algorithm, the Lie transform method is triggered only when $\epsilon$ dips below a predetermined value.  One then transforms the variables, evolves the new Hamiltonian, and waits for $\epsilon$ to go above some de-triggering value before transforming back.  It is fruitful to implement this algorithm in an object oriented manner such that the collection of N vortices is itself an object.  When the method is triggered, we simply make a new (N-1)-vortex object with the transformed variables, and keep track of $J$ and $\theta$ externally.  This view needs to be modified when you would like to use the transformation at fourth order or higher, where the modification to the Hamiltonian forces us to integrate the internal vortex dimer variables in conjunction with all the other variables.  If using the Lie transforms only to third order does not seem to be much of a sacrifice, then the N-vortex object allows us to deal with more than one dimer pair at the same time.  In particular we would just ``pop" down one more level to an (N-2)-vortex system where the integrable dimer motions are treated separately.  In particular, this would allow us to treat our simplest case of three vortices in a completely integrable fashion, where after considering the close vortices as a dimer, we could then look at the dimer and remaining vortex as a dimer itself.  In this way, our method finds subsections of the phase-space that are integrable in a particular manner and exploits that for numeric gain.  So, one must choose at third order whether one wants increased accuracy or the ability to treat other vortex pairs in a similar manner.

\section{Summary and Conclusions}

The 2D point-vortex model is a very useful, albeit simple, approximation to many important physical systems.  The forward time dynamics are easily obtained from a coupled nonlinear system of ODEs.  However several problems arise when using an adaptive time-stepping method to numerically integrate the system while two like-signed vortices are very close compared to other vortex separations.  If one keeps a small tolerance, the time-step will become prohibitively small and unnecessarily slow down the integration of the whole system.  On the other hand, if one increases the tolerance to alleviate this problem, not only does the accuracy suffer, but the overall energy of the supposedly Hamiltonian system tends to drift. \\

	To address these issues, we transformed the system such that it separated into two separate components: the motion of the vortex pair, or dimer, about its center of circulation and the motion of the dimer itself, considered as a single vortex, along with the rest of the vortices.  We viewed this as an (N-1)-vortex system with the same dynamics, where one of the vortices, the dimer, has an internal ``spin" degree of freedom.  The corrections to this appealing picture were fortunately ordered in a small parameter.  We then used Lie transform perturbation theory to develop a near identity canonical transformation that preserved this picture order by order in the small parameter.  At fourth order, a slight alteration to the dynamics was necessary, though this affected only the dimer's spin degree of freedom by slightly lowering its rotation rate.  With this inclusion, the (N-1)-vortex picture was altered only in so far as the spin of the dimer weakly coupled to a new field produced by the (N-1) vortices.  \\

	When quantifying the efficacy of the algorithmic implementation of our transformations, the Lie method, we chose the particularly simple example of three vortices of equal circulation with one close pair.   The first metric we considered was the speed-up factor, which measures the ratio of the number of calls to the numerical integrator made by the regular method versus those made by the Lie method.  Most importantly, we found this value to scale as $\epsilon^{-1.8}$, which indicates that the regular method run-time diverges as $\epsilon$ becomes small, since the Lie method run-time remains constant.  Next we considered the accuracy of the Lie method versus the real motion as given by the regular method with a very low tolerance.  The accuracy is quantified by the distance between two initially identical phase space points after evolving under the regular and Lie methods for a unit time.  This was further broken down into the the phase space variables of the dimer relative to the center of circulation, and of the center of circulation and remaining vortex.  For both cases we found that the measure of accuracy gets smaller, i.e. more accurate, as a $\epsilon$ decreases.  As expected, the relationship is a power-law, with the integer power depending on the Lie transform order that is used.  Additionally we examined the total energy of the system.  The Lie method has zero drift in the value of the Hamiltonian as well as very small amplitude oscillations.  These oscillations drastically decrease in amplitude as $\epsilon$ decreases, and do so in a manner that gets better with inclusion of successively higher orders of the Lie transformation.  This is to be contrasted with the consistent decrease in energy that occurs with the regular method when using an insufficiently low tolerance.  Finally, we described how this transformation could be used as the basis of an object-oriented method for accurate numerical integration of point-vortex dynamics.  \\

\bibliographystyle{apsrev}
\bibliography{VLTbib}

\end{document}

%% file: Efficiency.tex
\begingroup
  \makeatletter
  \providecommand\color[2][]{%
    \GenericError{(gnuplot) \space\space\space\@spaces}{%
      Package color not loaded in conjunction with
      terminal option `colourtext'%
    }{See the gnuplot documentation for explanation.%
    }{Either use 'blacktext' in gnuplot or load the package
      color.sty in LaTeX.}%
    \renewcommand\color[2][]{}%
  }%
  \providecommand\includegraphics[2][]{%
    \GenericError{(gnuplot) \space\space\space\@spaces}{%
      Package graphicx or graphics not loaded%
    }{See the gnuplot documentation for explanation.%
    }{The gnuplot epslatex terminal needs graphicx.sty or graphics.sty.}%
    \renewcommand\includegraphics[2][]{}%
  }%
  \providecommand\rotatebox[2]{#2}%
  \@ifundefined{ifGPcolor}{%
    \newif\ifGPcolor
    \GPcolorfalse
  }{}%
  \@ifundefined{ifGPblacktext}{%
    \newif\ifGPblacktext
    \GPblacktexttrue
  }{}%
  \let\gplgaddtomacro\g@addto@macro
  \gdef\gplbacktext{}%
  \gdef\gplfronttext{}%
  \makeatother
  \ifGPblacktext
    \def\colorrgb#1{}%
    \def\colorgray#1{}%
  \else
    \ifGPcolor
      \def\colorrgb#1{\color[rgb]{#1}}%
      \def\colorgray#1{\color[gray]{#1}}%
      \expandafter\def\csname LTw\endcsname{\color{white}}%
      \expandafter\def\csname LTb\endcsname{\color{black}}%
      \expandafter\def\csname LTa\endcsname{\color{black}}%
      \expandafter\def\csname LT0\endcsname{\color[rgb]{1,0,0}}%
      \expandafter\def\csname LT1\endcsname{\color[rgb]{0,1,0}}%
      \expandafter\def\csname LT2\endcsname{\color[rgb]{0,0,1}}%
      \expandafter\def\csname LT3\endcsname{\color[rgb]{1,0,1}}%
      \expandafter\def\csname LT4\endcsname{\color[rgb]{0,1,1}}%
      \expandafter\def\csname LT5\endcsname{\color[rgb]{1,1,0}}%
      \expandafter\def\csname LT6\endcsname{\color[rgb]{0,0,0}}%
      \expandafter\def\csname LT7\endcsname{\color[rgb]{1,0.3,0}}%
      \expandafter\def\csname LT8\endcsname{\color[rgb]{0.5,0.5,0.5}}%
    \else
      \def\colorrgb#1{\color{black}}%
      \def\colorgray#1{\color[gray]{#1}}%
      \expandafter\def\csname LTw\endcsname{\color{white}}%
      \expandafter\def\csname LTb\endcsname{\color{black}}%
      \expandafter\def\csname LTa\endcsname{\color{black}}%
      \expandafter\def\csname LT0\endcsname{\color{black}}%
      \expandafter\def\csname LT1\endcsname{\color{black}}%
      \expandafter\def\csname LT2\endcsname{\color{black}}%
      \expandafter\def\csname LT3\endcsname{\color{black}}%
      \expandafter\def\csname LT4\endcsname{\color{black}}%
      \expandafter\def\csname LT5\endcsname{\color{black}}%
      \expandafter\def\csname LT6\endcsname{\color{black}}%
      \expandafter\def\csname LT7\endcsname{\color{black}}%
      \expandafter\def\csname LT8\endcsname{\color{black}}%
    \fi
  \fi
  \setlength{\unitlength}{0.0500bp}%
  \begin{picture}(7200.00,5040.00)%
    \gplgaddtomacro\gplbacktext{%
      \csname LTb\endcsname%
      \put(990,660){\makebox(0,0)[r]{\strut{} 0}}%
      \put(990,1125){\makebox(0,0)[r]{\strut{} 50}}%
      \put(990,1590){\makebox(0,0)[r]{\strut{} 100}}%
      \put(990,2055){\makebox(0,0)[r]{\strut{} 150}}%
      \put(990,2520){\makebox(0,0)[r]{\strut{} 200}}%
      \put(990,2985){\makebox(0,0)[r]{\strut{} 250}}%
      \put(990,3450){\makebox(0,0)[r]{\strut{} 300}}%
      \put(990,3915){\makebox(0,0)[r]{\strut{} 350}}%
      \put(990,4380){\makebox(0,0)[r]{\strut{} 400}}%
      \put(1122,440){\makebox(0,0){\strut{} 0}}%
      \put(2073,440){\makebox(0,0){\strut{} 0.05}}%
      \put(3023,440){\makebox(0,0){\strut{} 0.1}}%
      \put(3974,440){\makebox(0,0){\strut{} 0.15}}%
      \put(4925,440){\makebox(0,0){\strut{} 0.2}}%
      \put(5875,440){\makebox(0,0){\strut{} 0.25}}%
      \put(6826,440){\makebox(0,0){\strut{} 0.3}}%
      \put(220,2520){\rotatebox{90}{\makebox(0,0){\strut{}$\frac{N_R}{N_L}$}}}%
      \put(3974,110){\makebox(0,0){\strut{}$\epsilon$}}%
      \put(3974,4710){\makebox(0,0){\strut{}Speedup and Efficiency}}%
    }%
    \gplgaddtomacro\gplfronttext{%
    }%
    \gplgaddtomacro\gplbacktext{%
      \csname LTb\endcsname%
      \put(3062,1952){\makebox(0,0)[r]{\strut{} 1}}%
      \put(3062,2692){\makebox(0,0)[r]{\strut{} 10}}%
      \put(3062,3431){\makebox(0,0)[r]{\strut{} 100}}%
      \put(3062,4171){\makebox(0,0)[r]{\strut{} 1000}}%
      \put(3194,1732){\makebox(0,0){\strut{} 0.01}}%
      \put(4830,1732){\makebox(0,0){\strut{} 0.1}}%
      \put(6466,1732){\makebox(0,0){\strut{} 1}}%
    }%
    \gplgaddtomacro\gplfronttext{%
    }%
    \gplbacktext
    \put(0,0){\includegraphics{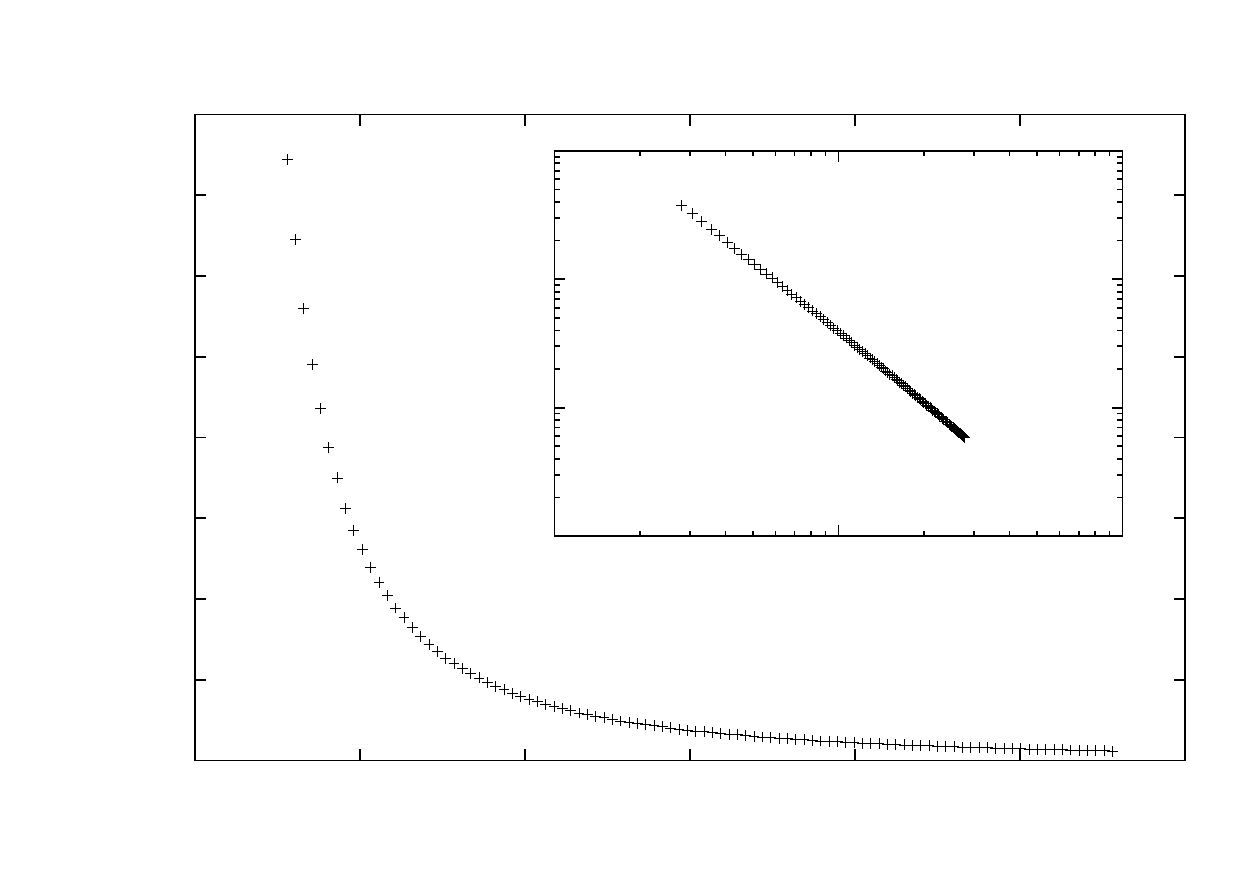}}%
    \gplfronttext
  \end{picture}%
\endgroup

%% file: AccuracyD.tex
\begingroup
  \makeatletter
  \providecommand\color[2][]{%
    \GenericError{(gnuplot) \space\space\space\@spaces}{%
      Package color not loaded in conjunction with
      terminal option `colourtext'%
    }{See the gnuplot documentation for explanation.%
    }{Either use 'blacktext' in gnuplot or load the package
      color.sty in LaTeX.}%
    \renewcommand\color[2][]{}%
  }%
  \providecommand\includegraphics[2][]{%
    \GenericError{(gnuplot) \space\space\space\@spaces}{%
      Package graphicx or graphics not loaded%
    }{See the gnuplot documentation for explanation.%
    }{The gnuplot epslatex terminal needs graphicx.sty or graphics.sty.}%
    \renewcommand\includegraphics[2][]{}%
  }%
  \providecommand\rotatebox[2]{#2}%
  \@ifundefined{ifGPcolor}{%
    \newif\ifGPcolor
    \GPcolorfalse
  }{}%
  \@ifundefined{ifGPblacktext}{%
    \newif\ifGPblacktext
    \GPblacktexttrue
  }{}%
  \let\gplgaddtomacro\g@addto@macro
  \gdef\gplbacktext{}%
  \gdef\gplfronttext{}%
  \makeatother
  \ifGPblacktext
    \def\colorrgb#1{}%
    \def\colorgray#1{}%
  \else
    \ifGPcolor
      \def\colorrgb#1{\color[rgb]{#1}}%
      \def\colorgray#1{\color[gray]{#1}}%
      \expandafter\def\csname LTw\endcsname{\color{white}}%
      \expandafter\def\csname LTb\endcsname{\color{black}}%
      \expandafter\def\csname LTa\endcsname{\color{black}}%
      \expandafter\def\csname LT0\endcsname{\color[rgb]{1,0,0}}%
      \expandafter\def\csname LT1\endcsname{\color[rgb]{0,1,0}}%
      \expandafter\def\csname LT2\endcsname{\color[rgb]{0,0,1}}%
      \expandafter\def\csname LT3\endcsname{\color[rgb]{1,0,1}}%
      \expandafter\def\csname LT4\endcsname{\color[rgb]{0,1,1}}%
      \expandafter\def\csname LT5\endcsname{\color[rgb]{1,1,0}}%
      \expandafter\def\csname LT6\endcsname{\color[rgb]{0,0,0}}%
      \expandafter\def\csname LT7\endcsname{\color[rgb]{1,0.3,0}}%
      \expandafter\def\csname LT8\endcsname{\color[rgb]{0.5,0.5,0.5}}%
    \else
      \def\colorrgb#1{\color{black}}%
      \def\colorgray#1{\color[gray]{#1}}%
      \expandafter\def\csname LTw\endcsname{\color{white}}%
      \expandafter\def\csname LTb\endcsname{\color{black}}%
      \expandafter\def\csname LTa\endcsname{\color{black}}%
      \expandafter\def\csname LT0\endcsname{\color{black}}%
      \expandafter\def\csname LT1\endcsname{\color{black}}%
      \expandafter\def\csname LT2\endcsname{\color{black}}%
      \expandafter\def\csname LT3\endcsname{\color{black}}%
      \expandafter\def\csname LT4\endcsname{\color{black}}%
      \expandafter\def\csname LT5\endcsname{\color{black}}%
      \expandafter\def\csname LT6\endcsname{\color{black}}%
      \expandafter\def\csname LT7\endcsname{\color{black}}%
      \expandafter\def\csname LT8\endcsname{\color{black}}%
    \fi
  \fi
  \setlength{\unitlength}{0.0500bp}%
  \begin{picture}(7200.00,5040.00)%
    \gplgaddtomacro\gplbacktext{%
      \csname LTb\endcsname%
      \put(1254,660){\makebox(0,0)[r]{\strut{} 0}}%
      \put(1254,1073){\makebox(0,0)[r]{\strut{} 0.005}}%
      \put(1254,1487){\makebox(0,0)[r]{\strut{} 0.01}}%
      \put(1254,1900){\makebox(0,0)[r]{\strut{} 0.015}}%
      \put(1254,2313){\makebox(0,0)[r]{\strut{} 0.02}}%
      \put(1254,2727){\makebox(0,0)[r]{\strut{} 0.025}}%
      \put(1254,3140){\makebox(0,0)[r]{\strut{} 0.03}}%
      \put(1254,3553){\makebox(0,0)[r]{\strut{} 0.035}}%
      \put(1254,3967){\makebox(0,0)[r]{\strut{} 0.04}}%
      \put(1254,4380){\makebox(0,0)[r]{\strut{} 0.045}}%
      \put(1386,440){\makebox(0,0){\strut{} 0}}%
      \put(2474,440){\makebox(0,0){\strut{} 0.05}}%
      \put(3562,440){\makebox(0,0){\strut{} 0.1}}%
      \put(4650,440){\makebox(0,0){\strut{} 0.15}}%
      \put(5738,440){\makebox(0,0){\strut{} 0.2}}%
      \put(6826,440){\makebox(0,0){\strut{} 0.25}}%
      \put(220,2520){\rotatebox{90}{\makebox(0,0){\strut{}$\frac{\left|\mbox{\bf\boldmath{z}}_{L,f}-\mbox{\bf\boldmath{z}}_{R,f}\right|}{\Delta t}$}}}%
      \put(4106,110){\makebox(0,0){\strut{}$\epsilon$}}%
      \put(4106,4710){\makebox(0,0){\strut{}Accuracy of the Dimer Internal Motion}}%
    }%
    \gplgaddtomacro\gplfronttext{%
    }%
    \gplgaddtomacro\gplbacktext{%
      \csname LTb\endcsname%
      \put(2030,2960){\makebox(0,0)[r]{\strut{} 1e-08}}%
      \put(2030,3147){\makebox(0,0)[r]{\strut{} 1e-07}}%
      \put(2030,3335){\makebox(0,0)[r]{\strut{} 1e-06}}%
      \put(2030,3522){\makebox(0,0)[r]{\strut{} 1e-05}}%
      \put(2030,3710){\makebox(0,0)[r]{\strut{} 0.0001}}%
      \put(2030,3897){\makebox(0,0)[r]{\strut{} 0.001}}%
      \put(2030,4085){\makebox(0,0)[r]{\strut{} 0.01}}%
      \put(2030,4272){\makebox(0,0)[r]{\strut{} 0.1}}%
      \put(2162,2740){\makebox(0,0){\strut{} 0.01}}%
      \put(3126,2740){\makebox(0,0){\strut{} 0.1}}%
      \put(4090,2740){\makebox(0,0){\strut{} 1}}%
    }%
    \gplgaddtomacro\gplfronttext{%
    }%
    \gplbacktext
    \put(0,0){\includegraphics{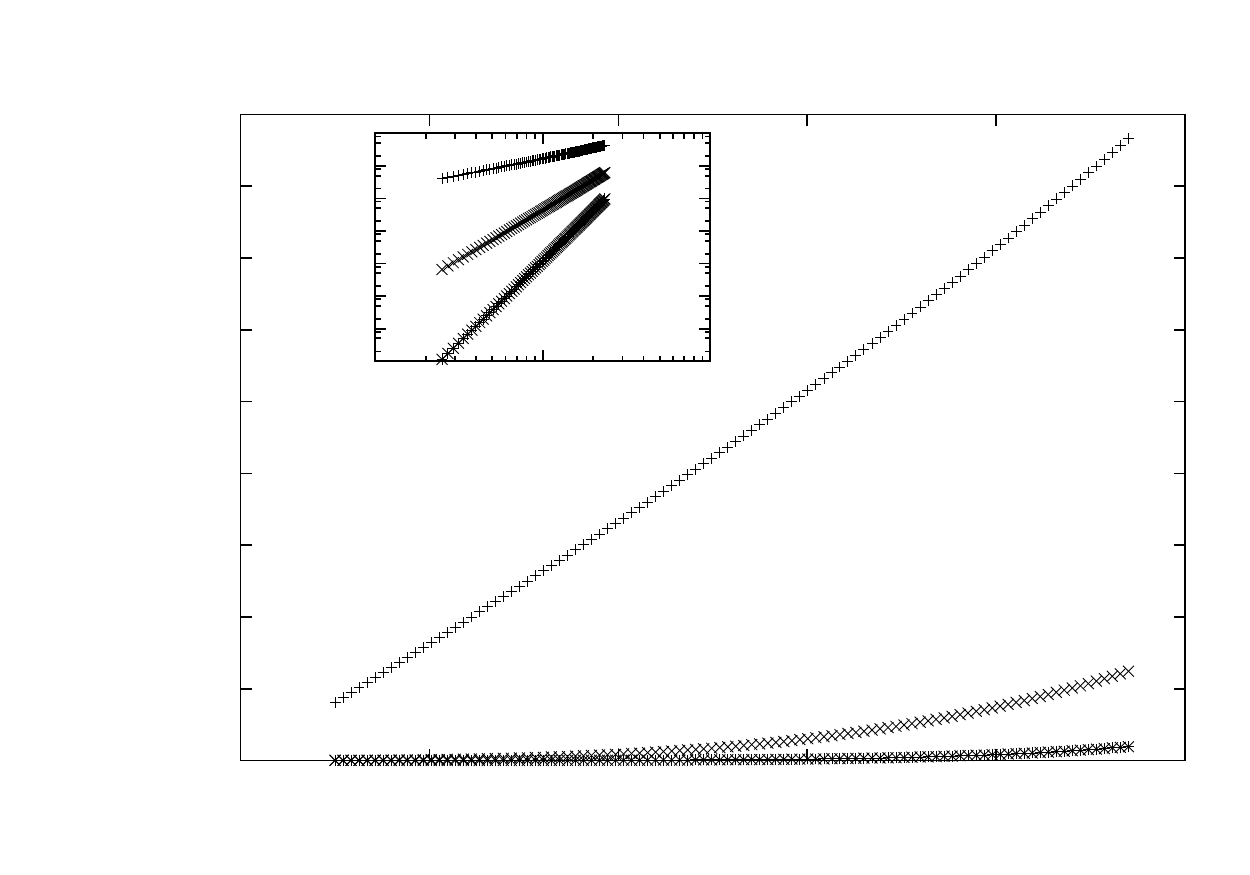}}%
    \gplfronttext
  \end{picture}%
\endgroup

%% file: AccuracyR.tex
\begingroup
  \makeatletter
  \providecommand\color[2][]{%
    \GenericError{(gnuplot) \space\space\space\@spaces}{%
      Package color not loaded in conjunction with
      terminal option `colourtext'%
    }{See the gnuplot documentation for explanation.%
    }{Either use 'blacktext' in gnuplot or load the package
      color.sty in LaTeX.}%
    \renewcommand\color[2][]{}%
  }%
  \providecommand\includegraphics[2][]{%
    \GenericError{(gnuplot) \space\space\space\@spaces}{%
      Package graphicx or graphics not loaded%
    }{See the gnuplot documentation for explanation.%
    }{The gnuplot epslatex terminal needs graphicx.sty or graphics.sty.}%
    \renewcommand\includegraphics[2][]{}%
  }%
  \providecommand\rotatebox[2]{#2}%
  \@ifundefined{ifGPcolor}{%
    \newif\ifGPcolor
    \GPcolorfalse
  }{}%
  \@ifundefined{ifGPblacktext}{%
    \newif\ifGPblacktext
    \GPblacktexttrue
  }{}%
  \let\gplgaddtomacro\g@addto@macro
  \gdef\gplbacktext{}%
  \gdef\gplfronttext{}%
  \makeatother
  \ifGPblacktext
    \def\colorrgb#1{}%
    \def\colorgray#1{}%
  \else
    \ifGPcolor
      \def\colorrgb#1{\color[rgb]{#1}}%
      \def\colorgray#1{\color[gray]{#1}}%
      \expandafter\def\csname LTw\endcsname{\color{white}}%
      \expandafter\def\csname LTb\endcsname{\color{black}}%
      \expandafter\def\csname LTa\endcsname{\color{black}}%
      \expandafter\def\csname LT0\endcsname{\color[rgb]{1,0,0}}%
      \expandafter\def\csname LT1\endcsname{\color[rgb]{0,1,0}}%
      \expandafter\def\csname LT2\endcsname{\color[rgb]{0,0,1}}%
      \expandafter\def\csname LT3\endcsname{\color[rgb]{1,0,1}}%
      \expandafter\def\csname LT4\endcsname{\color[rgb]{0,1,1}}%
      \expandafter\def\csname LT5\endcsname{\color[rgb]{1,1,0}}%
      \expandafter\def\csname LT6\endcsname{\color[rgb]{0,0,0}}%
      \expandafter\def\csname LT7\endcsname{\color[rgb]{1,0.3,0}}%
      \expandafter\def\csname LT8\endcsname{\color[rgb]{0.5,0.5,0.5}}%
    \else
      \def\colorrgb#1{\color{black}}%
      \def\colorgray#1{\color[gray]{#1}}%
      \expandafter\def\csname LTw\endcsname{\color{white}}%
      \expandafter\def\csname LTb\endcsname{\color{black}}%
      \expandafter\def\csname LTa\endcsname{\color{black}}%
      \expandafter\def\csname LT0\endcsname{\color{black}}%
      \expandafter\def\csname LT1\endcsname{\color{black}}%
      \expandafter\def\csname LT2\endcsname{\color{black}}%
      \expandafter\def\csname LT3\endcsname{\color{black}}%
      \expandafter\def\csname LT4\endcsname{\color{black}}%
      \expandafter\def\csname LT5\endcsname{\color{black}}%
      \expandafter\def\csname LT6\endcsname{\color{black}}%
      \expandafter\def\csname LT7\endcsname{\color{black}}%
      \expandafter\def\csname LT8\endcsname{\color{black}}%
    \fi
  \fi
  \setlength{\unitlength}{0.0500bp}%
  \begin{picture}(7200.00,5040.00)%
    \gplgaddtomacro\gplbacktext{%
      \csname LTb\endcsname%
      \put(1386,660){\makebox(0,0)[r]{\strut{} 0}}%
      \put(1386,1280){\makebox(0,0)[r]{\strut{} 0.0005}}%
      \put(1386,1900){\makebox(0,0)[r]{\strut{} 0.001}}%
      \put(1386,2520){\makebox(0,0)[r]{\strut{} 0.0015}}%
      \put(1386,3140){\makebox(0,0)[r]{\strut{} 0.002}}%
      \put(1386,3760){\makebox(0,0)[r]{\strut{} 0.0025}}%
      \put(1386,4380){\makebox(0,0)[r]{\strut{} 0.003}}%
      \put(1518,440){\makebox(0,0){\strut{} 0}}%
      \put(2580,440){\makebox(0,0){\strut{} 0.05}}%
      \put(3641,440){\makebox(0,0){\strut{} 0.1}}%
      \put(4703,440){\makebox(0,0){\strut{} 0.15}}%
      \put(5764,440){\makebox(0,0){\strut{} 0.2}}%
      \put(6826,440){\makebox(0,0){\strut{} 0.25}}%
      \put(220,2520){\rotatebox{90}{\makebox(0,0){\strut{}$\frac{\left|\mbox{\bf\boldmath{z}}_{L,f}-\mbox{\bf\boldmath{z}}_{R,f}\right|}{\Delta t}$}}}%
      \put(4172,110){\makebox(0,0){\strut{}$\epsilon$}}%
      \put(4172,4710){\makebox(0,0){\strut{}Accuracy of the Rest of the System}}%
    }%
    \gplgaddtomacro\gplfronttext{%
    }%
    \gplgaddtomacro\gplbacktext{%
      \csname LTb\endcsname%
      \put(2174,2406){\makebox(0,0)[r]{\strut{} 1e-08}}%
      \put(2174,2709){\makebox(0,0)[r]{\strut{} 1e-07}}%
      \put(2174,3011){\makebox(0,0)[r]{\strut{} 1e-06}}%
      \put(2174,3314){\makebox(0,0)[r]{\strut{} 1e-05}}%
      \put(2174,3617){\makebox(0,0)[r]{\strut{} 0.0001}}%
      \put(2174,3919){\makebox(0,0)[r]{\strut{} 0.001}}%
      \put(2174,4222){\makebox(0,0)[r]{\strut{} 0.01}}%
      \put(2306,2186){\makebox(0,0){\strut{} 0.01}}%
      \put(3630,2186){\makebox(0,0){\strut{} 0.1}}%
      \put(4954,2186){\makebox(0,0){\strut{} 1}}%
    }%
    \gplgaddtomacro\gplfronttext{%
    }%
    \gplbacktext
    \put(0,0){\includegraphics{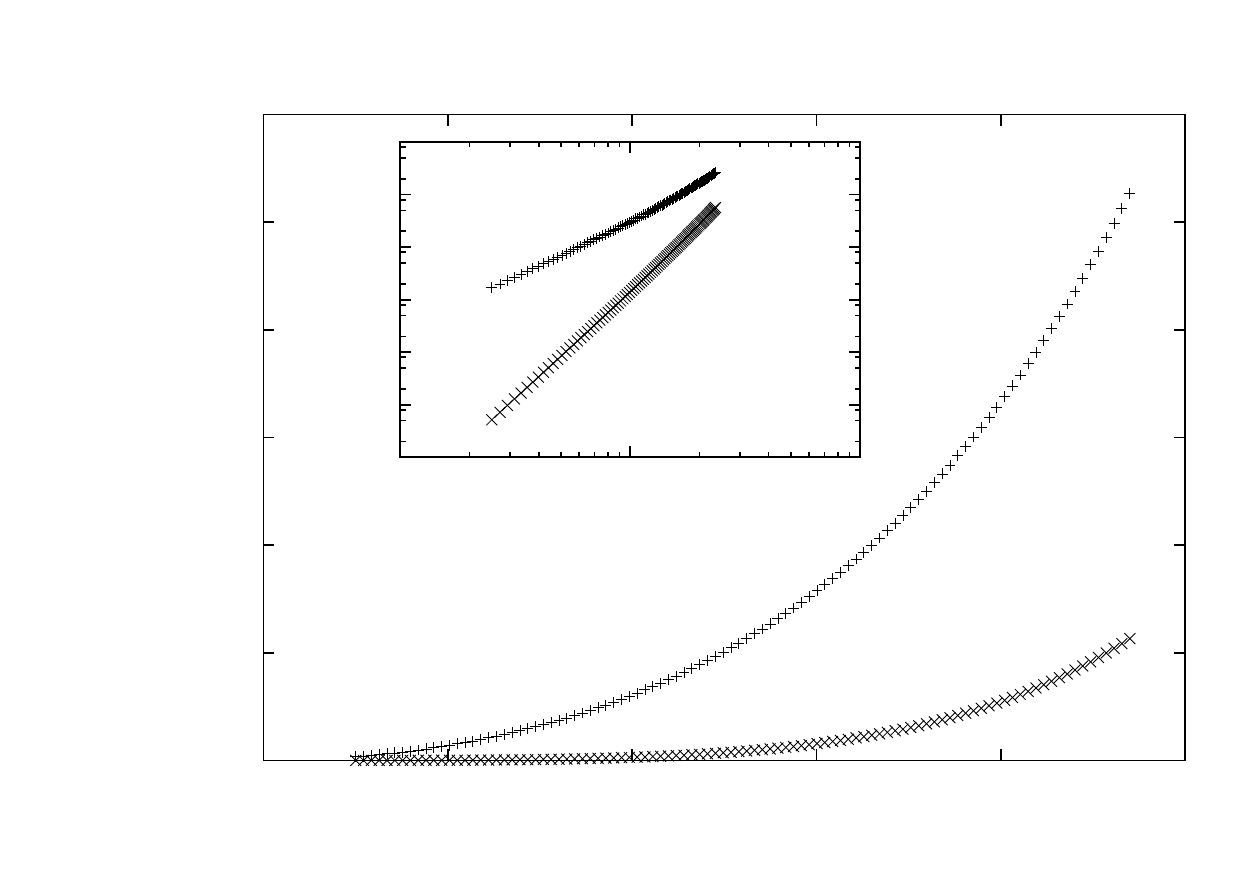}}%
    \gplfronttext
  \end{picture}%
\endgroup

%% file: Energy.tex
\begingroup
  \makeatletter
  \providecommand\color[2][]{%
    \GenericError{(gnuplot) \space\space\space\@spaces}{%
      Package color not loaded in conjunction with
      terminal option `colourtext'%
    }{See the gnuplot documentation for explanation.%
    }{Either use 'blacktext' in gnuplot or load the package
      color.sty in LaTeX.}%
    \renewcommand\color[2][]{}%
  }%
  \providecommand\includegraphics[2][]{%
    \GenericError{(gnuplot) \space\space\space\@spaces}{%
      Package graphicx or graphics not loaded%
    }{See the gnuplot documentation for explanation.%
    }{The gnuplot epslatex terminal needs graphicx.sty or graphics.sty.}%
    \renewcommand\includegraphics[2][]{}%
  }%
  \providecommand\rotatebox[2]{#2}%
  \@ifundefined{ifGPcolor}{%
    \newif\ifGPcolor
    \GPcolorfalse
  }{}%
  \@ifundefined{ifGPblacktext}{%
    \newif\ifGPblacktext
    \GPblacktexttrue
  }{}%
  \let\gplgaddtomacro\g@addto@macro
  \gdef\gplbacktext{}%
  \gdef\gplfronttext{}%
  \makeatother
  \ifGPblacktext
    \def\colorrgb#1{}%
    \def\colorgray#1{}%
  \else
    \ifGPcolor
      \def\colorrgb#1{\color[rgb]{#1}}%
      \def\colorgray#1{\color[gray]{#1}}%
      \expandafter\def\csname LTw\endcsname{\color{white}}%
      \expandafter\def\csname LTb\endcsname{\color{black}}%
      \expandafter\def\csname LTa\endcsname{\color{black}}%
      \expandafter\def\csname LT0\endcsname{\color[rgb]{1,0,0}}%
      \expandafter\def\csname LT1\endcsname{\color[rgb]{0,1,0}}%
      \expandafter\def\csname LT2\endcsname{\color[rgb]{0,0,1}}%
      \expandafter\def\csname LT3\endcsname{\color[rgb]{1,0,1}}%
      \expandafter\def\csname LT4\endcsname{\color[rgb]{0,1,1}}%
      \expandafter\def\csname LT5\endcsname{\color[rgb]{1,1,0}}%
      \expandafter\def\csname LT6\endcsname{\color[rgb]{0,0,0}}%
      \expandafter\def\csname LT7\endcsname{\color[rgb]{1,0.3,0}}%
      \expandafter\def\csname LT8\endcsname{\color[rgb]{0.5,0.5,0.5}}%
    \else
      \def\colorrgb#1{\color{black}}%
      \def\colorgray#1{\color[gray]{#1}}%
      \expandafter\def\csname LTw\endcsname{\color{white}}%
      \expandafter\def\csname LTb\endcsname{\color{black}}%
      \expandafter\def\csname LTa\endcsname{\color{black}}%
      \expandafter\def\csname LT0\endcsname{\color{black}}%
      \expandafter\def\csname LT1\endcsname{\color{black}}%
      \expandafter\def\csname LT2\endcsname{\color{black}}%
      \expandafter\def\csname LT3\endcsname{\color{black}}%
      \expandafter\def\csname LT4\endcsname{\color{black}}%
      \expandafter\def\csname LT5\endcsname{\color{black}}%
      \expandafter\def\csname LT6\endcsname{\color{black}}%
      \expandafter\def\csname LT7\endcsname{\color{black}}%
      \expandafter\def\csname LT8\endcsname{\color{black}}%
    \fi
  \fi
  \setlength{\unitlength}{0.0500bp}%
  \begin{picture}(7200.00,5040.00)%
    \gplgaddtomacro\gplbacktext{%
      \csname LTb\endcsname%
      \put(1386,660){\makebox(0,0)[r]{\strut{} 1e-11}}%
      \put(1386,1073){\makebox(0,0)[r]{\strut{} 1e-10}}%
      \put(1386,1487){\makebox(0,0)[r]{\strut{} 1e-09}}%
      \put(1386,1900){\makebox(0,0)[r]{\strut{} 1e-08}}%
      \put(1386,2313){\makebox(0,0)[r]{\strut{} 1e-07}}%
      \put(1386,2727){\makebox(0,0)[r]{\strut{} 1e-06}}%
      \put(1386,3140){\makebox(0,0)[r]{\strut{} 1e-05}}%
      \put(1386,3553){\makebox(0,0)[r]{\strut{} 0.0001}}%
      \put(1386,3967){\makebox(0,0)[r]{\strut{} 0.001}}%
      \put(1386,4380){\makebox(0,0)[r]{\strut{} 0.01}}%
      \put(1518,440){\makebox(0,0){\strut{} 0.02}}%
      \put(2877,440){\makebox(0,0){\strut{} 0.04}}%
      \put(3671,440){\makebox(0,0){\strut{} 0.06}}%
      \put(4235,440){\makebox(0,0){\strut{} 0.08}}%
      \put(4673,440){\makebox(0,0){\strut{} 0.1}}%
      \put(6031,440){\makebox(0,0){\strut{} 0.2}}%
      \put(6826,440){\makebox(0,0){\strut{} 0.3}}%
      \put(220,2520){\rotatebox{90}{\makebox(0,0){\strut{}$A \; \; \& \; \; \frac{dH}{dt}$}}}%
      \put(4172,110){\makebox(0,0){\strut{}$\epsilon$}}%
      \put(4172,4710){\makebox(0,0){\strut{}Energy Conservation}}%
    }%
    \gplgaddtomacro\gplfronttext{%
    }%
    \gplbacktext
    \put(0,0){\includegraphics{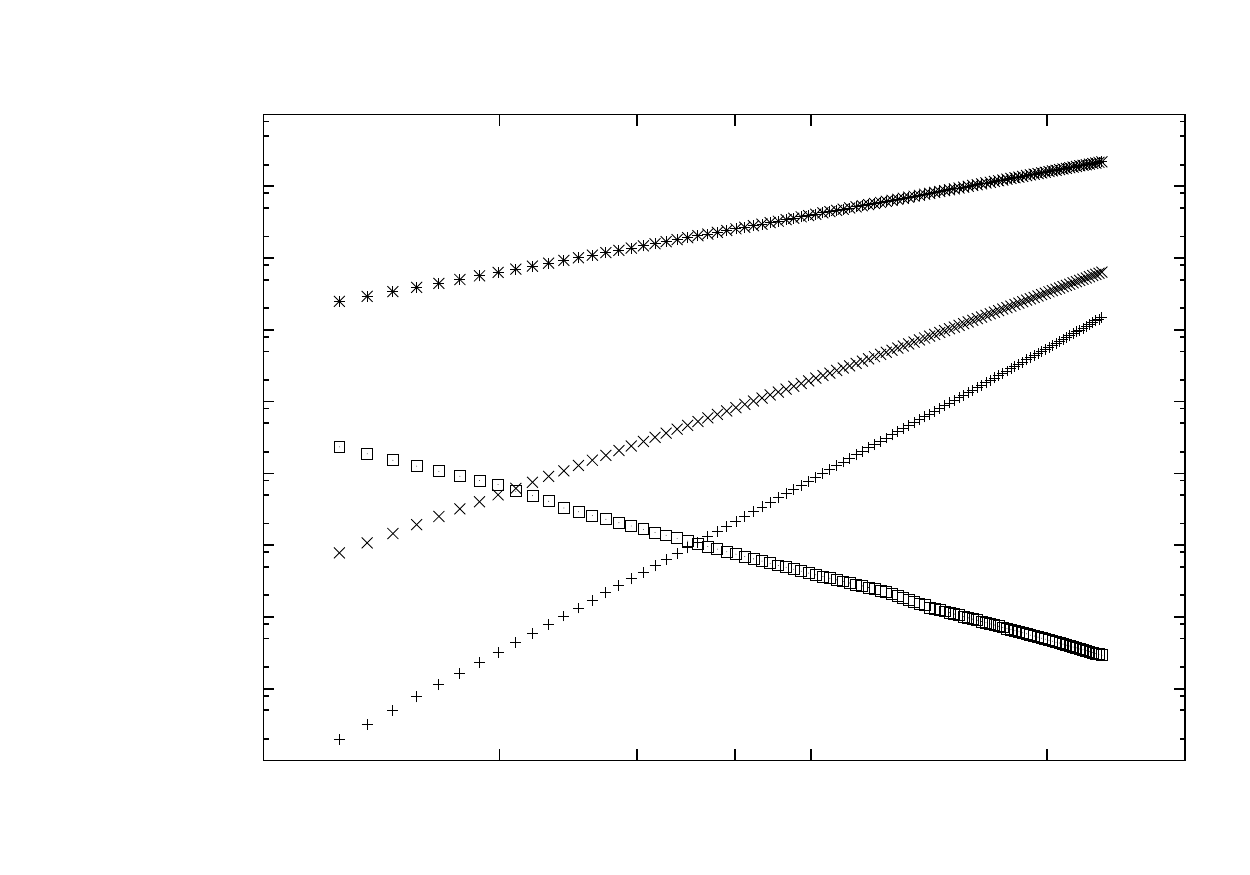}}%
    \gplfronttext
  \end{picture}%
\endgroup